\documentclass[12pt]{article}

\usepackage{scicite}
\usepackage{amssymb}
\usepackage{graphicx}
\usepackage{subcaption}
\usepackage{url}
\usepackage{bm}
\usepackage{xcolor}
\usepackage{aas_macros}
\usepackage{placeins}

\usepackage{times}
\usepackage{chemformula}
\usepackage{textcomp}
\usepackage{xspace}

\newcommand{\HI}{\hbox{{\rm H}{\sc \,i}}}
\newcommand{\um}{\textmu m\xspace}
\usepackage{booktabs} 
\usepackage{dcolumn}
\newcolumntype{d}[1]{D{.}{.}{#1}}

\usepackage{xcolor}
\usepackage{soul}
\definecolor{darkorange}{rgb}{1.0, 0.55, 0.0}

\newenvironment{sciabstract}{%
\begin{quote} \bf}
{\end{quote}}

\date{}

\begin{document} 
\begin{center}
{\LARGE Formation of the Methyl Cation by Photochemistry in a Protoplanetary Disk
}\\ 
\end{center}
Olivier Bern\'{e}$^1$,
Marie-Aline Martin-Drumel$^2$,
Ilane Schroetter$^1$,
Javier R. Goicoechea$^3$,
Ugo	Jacovella$^2$,
Bérenger Gans$^2$,
Emmanuel Dartois$^2$,
Laurent Coudert$^2$,
Edwin Bergin$^4$,
Felipe Alarcon$^4$,
Jan Cami$^{5 ,6, 7}$,
Evelyne Roueff$^{8}$,
John H.~Black$^{9}$,
Oskar Asvany$^{10}$,
Emilie Habart$^{11}$,
Els Peeters$^{5 ,6 , 7}$,
Amelie Canin$^{1}$,
Boris Trahin$^{11}$,
Christine Joblin$^1$,
Stephan Schlemmer$^{10}$,
Sven Thorwirth$^{10}$, 
Jose Cernicharo$^3$,
Maryvonne Gerin$^{8}$,
Alexander Tielens$^{12,13}$,
Marion Zannese$^{11}$,
Alain Abergel$^{11}$,
Jeronimo Bernard-Salas$^{14,15}$,
Christiaan Boersma$^{16}$,
Emeric Bron$^{8}$,
Ryan Chown$^{5, 6}$,
Sara Cuadrado$^{3}$,
Daniel Dicken$^{11}$,
Meriem Elyajouri$^{11}$,
Asunci\'on Fuente$^{17}$,
Karl D.\ Gordon$^{18}$,
Lina Issa$^{1}$,
Olga Kannavou$^{11}$,
Baria Khan$^{5}$,
Ozan Lacinbala$^{27}$,
David Languignon$^{8}$,
Romane Le~Gal$^{19,20}$,
Alexandros Maragkoudakis$^{16}$,
Raphael Meshaka$^{8}$,
Yoko Okada$^{10}$,
Takashi Onaka$^{21,22}$,
Sofia Pasquini$^{5}$,
Marc W. Pound$^{13}$,
Massimo Robberto$^{18}$,
Markus R\"ollig$^{23,24}$,
Bethany Schefter$^{5}$,
Thi\'ebaut Schirmer$^{11,25}$,
Ameek Sidhu$^{5, 6}$,
Benoit Tabone$^{11}$,
Dries Van De Putte$^{18}$,
S\'ilvia Vicente$^{26}$,
Mark G. Wolfire$^{13}$. 
\\
\\
\small{
$^{1}$Institut de Recherche en Astrophysique et Plan\'etologie, Universit\'e de Toulouse, CNRS, CNES, UPS, 9 Av. du colonel Roche, 31028 Toulouse Cedex 04, France\\
$^{2}$Universit\'e Paris-Saclay, CNRS, Institut des Sciences Mol\'eculaires d'Orsay, 91405 Orsay, France\\
$^{3}$Instituto de F\'{\i}sica Fundamental (CSIC). Calle Serrano 121-123, 28006, Madrid, Spain.\\
$^{4}$Department of Astronomy, University of Michigan, 1085 South University Avenue, Ann Arbor, MI 48109, USA\\
$^{5}$Department of Physics \& Astronomy, The University of Western Ontario, London ON N6A 3K7, Canada\\
$^{6}$Institute for Earth and Space Exploration, The University of Western Ontario, London ON N6A 3K7, Canada   \\
$^{7}$Carl Sagan Center, SETI Institute, 339 Bernardo Avenue, Suite 200, Mountain View, CA 94043, USA\\
$^{8}$LERMA, Observatoire de Paris, PSL University, Sorbonne Universit\'e, CNRS, 5 place Janssen, 92190, Meudon Cedex, France.\\
$^{9}$Department of Space, Earth, and Environment, Chalmers University of Technology, Onsala Space Observatory, 43992 Onsala, Sweden.\\
$^{10}$I.  Physikalisches Institut, Universit\"at zu K\"oln, Z\"ulpicher Str.~77, 50937 K\"oln, Germany\\
$^{11}$Institut d'Astrophysique Spatiale, Universit\'e Paris-Saclay, CNRS,  B\^atiment 121, 91405 Orsay Cedex, France\\
$^{12}$Leiden Observatory, Leiden University, P.O. Box 9513, 2300 RA Leiden, The Netherlands\\
$^{13}$Astronomy Department, University of Maryland, College Park, MD 20742, USA\\
$^{14}$ACRI-ST, Centre d’Etudes et de Recherche de Grasse (CERGA), 10 Av. Nicolas Copernic, F-06130 Grasse, France\\
$^{15}$INCLASS Common Laboratory., 10 Av. Nicolas Copernic, 06130 Grasse, France\\
$^{16}$ NASA Ames Research Center, MS 245-6, Moffett Field, CA 94035-1000, USA\\
$^{17}$Observatorio Astron\'{o}mico Nacional (OAN,IGN), Alfonso XII, 3, E-28014 Madrid, Spain\\
$^{18}$Space Telescope Science Institute, 3700 San Martin Drive, Baltimore, MD 21218, USA\\
$^{19}$Institut de Plan\'etologie et d'Astrophysique de Grenoble (IPAG), Universit\'e Grenoble Alpes, CNRS, F-38000 Grenoble, France\\
$^{20}$Institut de Radioastronomie Millim\'etrique (IRAM), 300 Rue de la Piscine, F-38406 Saint-Martin d'H\`{e}res, France\\
$^{21}$Department of Physics, Faculty of Science and Engineering, Meisei University, 2-1-1 Hodokubo, Hino, Tokyo 191-8506, Japan\\
$^{22}$Department of Astronomy, Graduate School of Science, The University of Tokyo, 7-3-1 Bunkyo-ku, Tokyo 113-0033, Japan\\
$^{23}$ Physikalischer Verein - Gesellschaft f{\"u}r Bildung und Wissenschaft, Robert-Mayer-Str. 2, 60325 Frankfurt, Germany\\
$^{24}$ Goethe-Universit{\"a}t, Physikalisches Institut, Frankfurt am Main, Germany\\
$^{25}$Department of Space, Earth and Environment, Chalmers University of Technology, Onsala Space Observatory, SE-439 92 Onsala, Sweden\\
$^{26}$Instituto de Astrof\'isica e Ci\^{e}ncias do Espa\c co, Tapada da Ajuda, Edif\'icio Leste, 2\,$^{\circ}$ Piso, P-1349-018 Lisboa, Portugal\\
$^{27}$ KU Leuven Quantum Solid State Physics (QSP) Celestijnenlaan 200d - box 2414 3001 Leuven
}

\baselineskip24pt

\begin{sciabstract}

Forty years ago it was proposed that gas phase organic chemistry in the interstellar medium was initiated by the methyl cation  
\ch{CH3+} \cite{black1977models, smith1992ion, herbst2021unusual}, but hitherto it has not been observed outside the Solar System 
\cite{roueff2013ch2d+, indriolo2010constraining}. Alternative routes involving processes on grain surfaces have been invoked 
\cite{cuppen2017grain, semenov2010chemistry}. Here we report JWST observations of \ch{CH3+} in a protoplanetary disk in the 
Orion star forming region. We find that gas-phase organic chemistry is activated by UV irradiation.

\end{sciabstract}

\begin{figure*}[h!]
   \centering
   \resizebox{\hsize}{!}{\includegraphics[width=14cm]{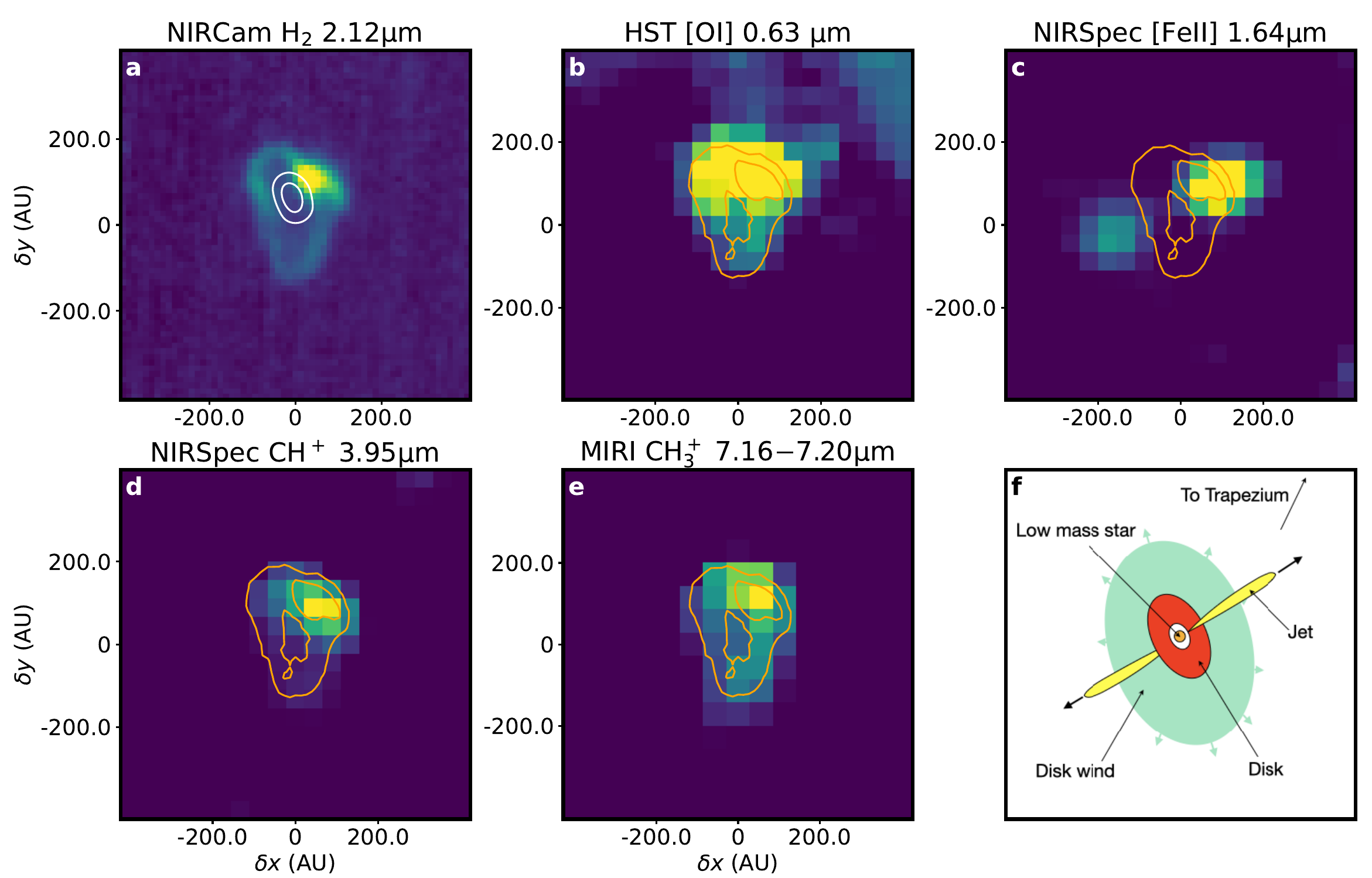}}
      \caption{ Overview of the d203-506 externally irradiated protoplanetary disk. Integrated intensity images from NIRCam F212N filter ({\bf a}), Hubble Space Telescope [OI] ({\bf b}),  NIRSpec [FeII] ({\bf c}), NIRSpec CH$^+$ 1-0 $P(7)$ at 3.95 \um ({\bf d}), and MIRI MRS integrated from 7.16 to 7.20 \um assigned to CH$_3^+$ ({\bf e}).
      Each panel is centered at $\alpha$=5:35:20.318 and $\delta$=-5:25:05.662 and are $2'' \times 2''$ wide. 
      Contours of NIRCam vibrationally excited H$_2$ (2.12 \um~band) are represented in orange, and ALMA dust continuum emission at 344 GHz from the disk in white contours. {({\bf f,}) Sketch of the d203-506 disk, jet, and wind. We note that the low mass star is not seen in the images because of the disk flaring. All JWST images are from the PDRs4All program, while the Hubble image is from \cite{bally2000disks}.  }
 }
         \label{fig:nb}
   \end{figure*}

As part of the PDRs4All Early Release Science program  on the JWST\footnote{pdrs4all.org} \cite{pdrs4all2022}, we have obtained observations of the protoplanetary disk d203-506 \cite{bally2000disks}. This object is situated in the Orion Bar, at about 0.25 pc from the massive, strongly UV emitting Trapezium  stars which are at 414 pc from Earth \cite{MentenK_07}, inside the Orion Nebula. The disk is about 100 au in radius, and has an estimated mass of $\sim 10 M_{ \rm Jup}$ (Bern\'e et al. in prep.). 
The central star of d203-506 has a estimated mass of $M_{\star}=0.2 \pm 0.1 M_{\odot}$ (Bern\'e et al. in prep.),
{typical for stars of the Orion Nebula Cluster \cite{hillenbrand2000constraints}}. This star is obscured by the flared disk that is seen nearly edge-on \cite{bally2000disks}. 
Fig.~\ref{fig:nb} shows integrated intensity images of the d203-506 disk (see Methods for details on JWST data reduction).
{
This includes the emission of vibrationally and rotationally excited H$_2$ and CH$^+$, and fine-structure emission of Oxygen ([OI]) and ionized iron ([FeII]). The molecular emission arises from a hot ($T_{\rm gas} \sim 1000\,$K) and dense ($n_\mathrm{H} > 10^{5}$ cm$^{-3}$) wind that is produced by photoevaporation from the disk due to irradiation by far-UV photons (FUV; 6\,$<$\,$E$\,$<$\,13.6~eV) from the Trapezium stars (Bern\'e et al. in prep.). The [FeII] image shows the emission associated with a collimated jet. Some of the wind emission is co-spatial with this jet, but overall the wind is more extended and creates a ``halo'' around the disk. }

\begin{figure*}[t]
   \centering
   \includegraphics[width=16cm]{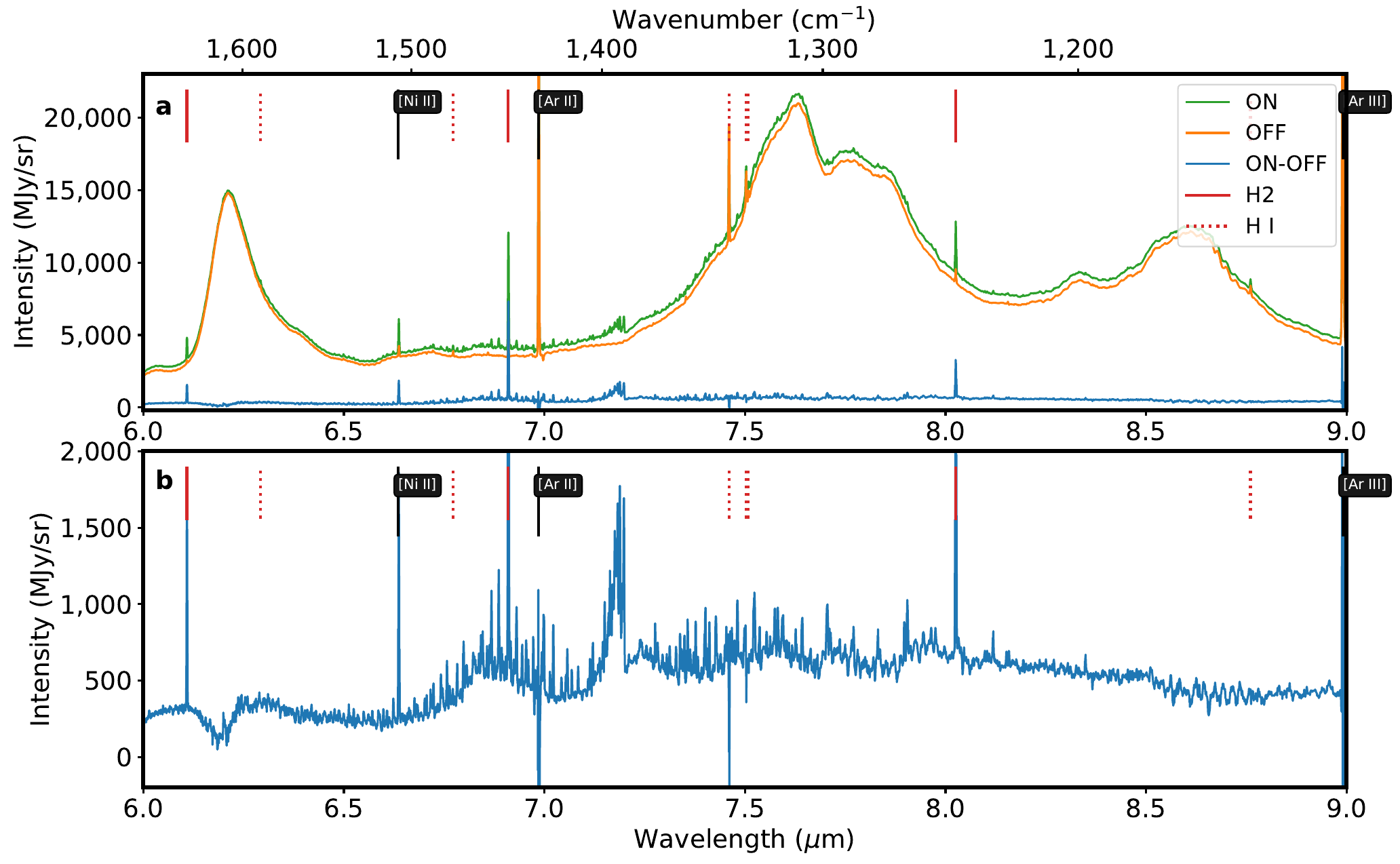}
      \caption{ JWST-MRIRI spectra of d203-506. 
      {\bf a}, Spectrum on the position of d203-506 (ON, green) and close to 203-506 (OFF, orange)
      over the 6-9 \um MIRI-MRS spectral range. The OFF spectrum is dominated by emission of the Orion Nebula:
       The broad features at 6.2, 7.7 and 8.6 \um are due to the emission of UV excited polycyclic aromatic hydrocarbons (PAHs, see Chown et al. in prep. for a detailed discussion).
      {\bf b}, Spectrum of d203-506 after subtraction of the nebular emission (ON $-$ OFF).      
              }
         \label{fig:spectrum}
   \end{figure*}

The mid infrared spectrum of d203-506 was obtained using the MIRI-MRS spectrometer onboard JWST (see Methods for details) and is shown in Fig.~\ref{fig:spectrum}. In the spectrum, we detect pure rotational lines of H$_2$ [0-0 S(1) to 0-0 S(8)] from which we derive an excitation temperature $T_{\rm ex} = 923 \pm 48 $ K (Methods).  %
The straight line observed in the excitation diagram derived from these lines (Extended Data Fig.~\ref{fig:h2rot}) indicates that  
the excitation temperature of H$_2$ is close to the gas kinetic temperature, and thus confirms the presence of hot molecular gas in the wind of d203-506.

In addition to identified H$_2$ and \HI~emission lines (see Methods), a strong residual emission consisting in a series of lines in the 6.5--8.0 \um range is observed (Fig.~\ref{fig:spectrum}).
This emission, as seen in the lower middle panel of  Fig.~\ref{fig:nb}, is spatially resolved and only present in d203-506. It is co-spatial with \ch{H_2} and CH$^+$ emission, {with the best spatial correlation observed with the 2.12\,\um line of vibrationally excited H$_2$ (see Fig.~\ref{fig:nb})}. 
We conclude that the observed features in Fig.~\ref{fig:spectrum} are an astrophysical signal associated with emission from the {wind of the d203-506 protoplanetary disk. We note however, that given that the angular resolution of MIRI is at the limit to resolve the structures in d203-506, we cannot fully exclude an emission contribution from the jet.}

\begin{figure}[ht!]
    \centering
    \includegraphics[width=\textwidth]{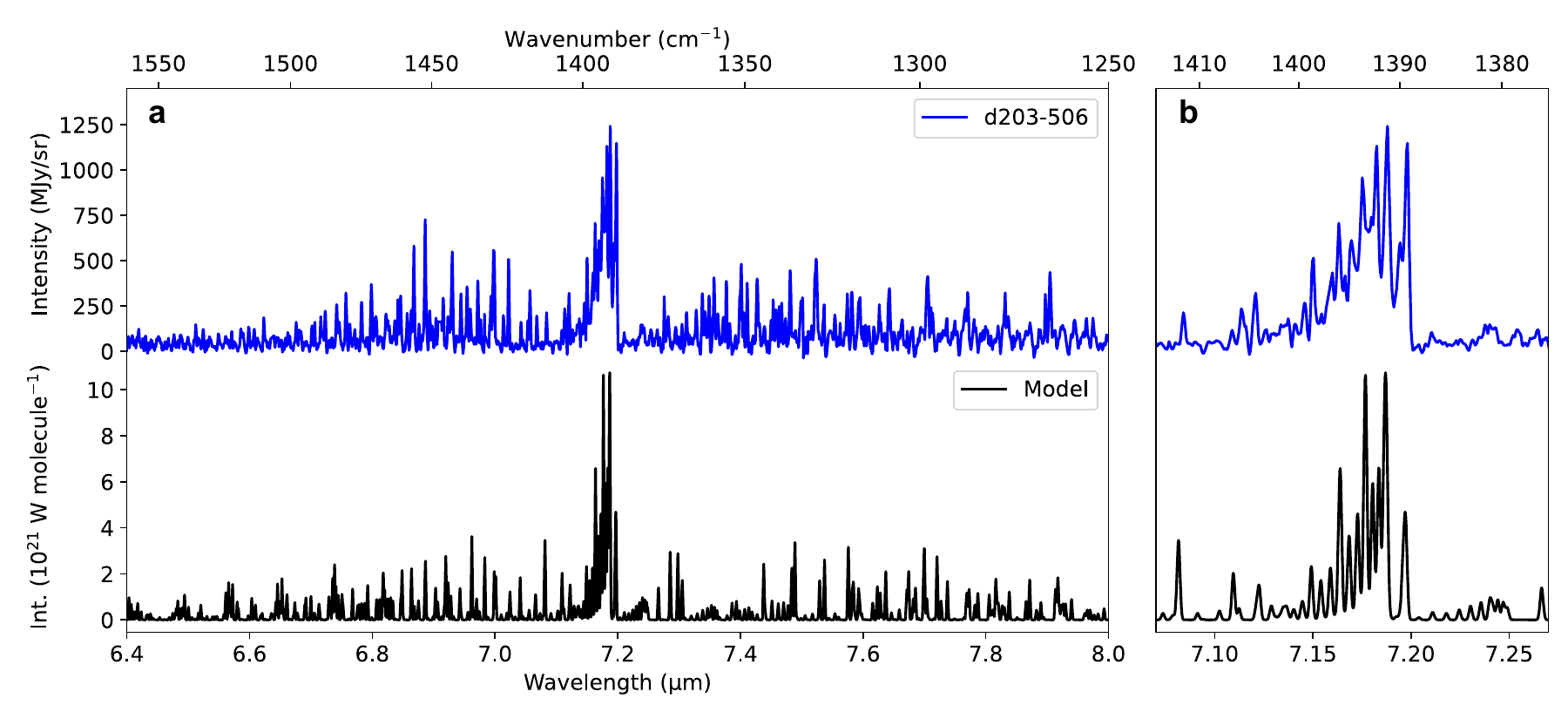}
    \caption{Comparison between the observed JWST spectrum of d203-506 and modeled CH$_3^+$ spectrum. 
    {\bf a,} Full spectrum. {\bf b,} Zoom on the strongest lines. 
    The model (black curve) for the $\nu_2$ and $\nu_4$ vibrations of CH$_3^+$ was obtained using the constants of Table~\ref{table:fit_results} (see Methods for a more detailed description of the simulation). A Gaussian linewidth of 0.35 cm$^{-1}$, corresponding to the MIRI MRS resolution in this range ($\frac{\lambda}{\Delta \lambda} \sim 3800$), has been used in the simulation performed using PGOPHER \cite{western2017:PGOPHER}. {For clarity, the observational spectrum has been baseline corrected using a spline function and strong individual lines  have been removed from the plot : [Ni II] at 6.63,
    H$_2$ at 6.92, [Ar II] at 6.99, He I at 7.47, and H$_2$ at 8.02~\um.  The standard deviation (1$\sigma$) of  the noise level is $\sim 10$ MJy~sr$^{-1}$ in this range. The observed lines are 10 to 100 times this noise level. }}
    \label{fig:mainmodelIII}
\end{figure}

The 7 \um band visible in Fig.~\ref{fig:spectrum} is composed of a succession of narrow features corresponding to ro-vibrational transitions of a molecular carrier.
Such insight into the detailed structure of the band is enabled by the unprecedented high spectral resolution and high sensitivity provided by JWST in that spectral region. 
The presence of these resolved structures, and their spectral span, is compatible
with a light molecular carrier.
The wavelength coincidence between the observed emission features around 7~\um (Fig.~\ref{fig:spectrum}) and the $\nu_2$ (out-of-plane bending, ``umbrella" motion) and $\nu_4$ (in-plane bending) bands of \ch{CH3+} \cite{CunhadeMiranda2010} is striking.  Besides \ch{CH3+}, not a single match has been found for a molecule that would possess its shortest wavelength emission signature at 7 \um 
(see details in the Supplementary information). 
{Recent laboratory work  on the low temperature vibrational spectroscopy of \ch{CH3+} \cite{asvany2018spectroscopy} finds the vibrational bands match the observed  wavelengths.}
Two additional spectroscopic analyses further strengthen the \ch{CH3+} assignment. 
First, the intensity pattern of successive emission lines (Fig.~\ref{fig:mainmodelIII}) is characteristic of the spin statistics of a molecular carrier suggesting three equivalent non-zero-spin atoms (e.g., hydrogen atoms), as expected for CH$_3^+$. 
Second, the observed emission spectrum can be satisfactorily simulated (see Fig.~\ref{fig:mainmodelIII}) using sets of spectroscopic constants taking values within the range of what is expected from available calculations (Refs. \cite{Kraemer1991:CH3+, Keceli2009:CH3+} and this work, see Table~\ref{table:fit_results})
{and laboratory measurements \cite{asvany2018spectroscopy}}. 
A detailed description of the spectroscopic analysis procedure is given in the Methods.
Quantum number assignments to the ro-vibrational transitions in this spectral range will require both additional theoretical input (notably investigating the Coriolis interaction between the two bands \cite{Kraemer1991:CH3+}) and laboratory measurements, in particular at even higher resolution and ideally at cold temperature to reduce the spectral density.
The higher energy $\nu_3$  band of CH$_3^+$, situated near 3~\um, has been measured 
at high resolution in the laboratory \cite{Crofton1988:CH3+nu3}. {Some of the expected lines from the $\nu_3$ band in this spectral range are tentatively detected with NIRSpec in d203-506 (see Supplementary Information and Extended Data Fig.~\ref{fig:NIRSpec}).}  
In summary, by spectroscopic standards, \ch{CH3+} is the best candidate to explain the 7 \um spectral band observed towards the d203-506 protoplanetary disk.

\begin{table}[h!]
\centering
\caption{Spectroscopic parameters of \ch{CH3+} in the two excited states  $\nu_2$ and $\nu_4$  from experiment and theory, and comparison with constants from our best model of the observed signatures. The ground state values ($v=0$) are kept fixed to the experimental values determined in Ref.~\cite{Crofton1988:CH3+nu3}. See Table \ref{tab:cts} for a more complete set of parameters.}
\begin{tabular}{cl cc c cc} 
\hline
 \multicolumn{1}{c}{Parameter}  & Unit & \multicolumn{2}{c}{$v_2=1$} && \multicolumn{2}{c}{$v_4=1$}\\ 
 &&  Prediction & Model && Prediction & Model\\
\hline
$\nu$ & cm$^{-1}$ &  1372--1412$^a$ & 1391 && 1373--1393$^b$ & 1375\\
 & \um & 7.289--7.082  & 7.133 && 7.283--7.179 & 7.273  \\
$B$ & cm$^{-1}$ & 9.06--9.49$^c$ & 9.37  && 9.48--9.52$^c$ & 9.50 \\
$C$ & cm$^{-1}$ & 4.61--4.66$^c$ & 4.66  && 4.55--4.65$^c$ & 4.57 \\
\hline
\label{table:fit_results}
\end{tabular}

\begin{minipage}{0.9\textwidth} \small
$^a$ 1$\sigma$ confidence interval from the experimental values of Ref.~\cite{asvany2018spectroscopy} (1402 $\pm$ 10 cm$^{-1}$) and Ref. \cite{CunhadeMiranda2010} (1387 $\pm$ 15 cm$^{-1}$). \\
$^b$ 1$\sigma$ confidence interval from the experimental value of Ref. \cite{asvany2018spectroscopy} (1383 $\pm$ 10 cm$^{-1}$)\\
$^c$  from Refs. \cite{Kraemer1991:CH3+, Keceli2009:CH3+} and this work, and scaled to the ground state parameters of  Ref.~\cite{Crofton1988:CH3+nu3}, see Table \ref{tab:cts}
\end{minipage}
\end{table}

The presence of CH$_{3}^{+}$ in d203-506 raises the question of its origin. 
Carbon chemistry in these environments typically starts by the  radiative association
\ch{C+ + H2 -> CH2+} + $h\nu$, which is a {very} slow process. The alternative bimolecular reaction
\ch{C+ + H2 \,($v$ = 0) -> CH+ + H}~[1], 
is endothermic by
$\Delta E$/$k_\mathrm{B}$\,=\,4300\,K  \cite{Hierl1997,Zanchet2013}, and thus very slow in cold ($T\sim 100$K) interstellar gas where \mbox{$T$\,$\ll$\,$\Delta E$/$k_\mathrm{B}$}. 
However, strong external FUV radiation fields combined with high gas densities as 
found in protoplanetary disks open new routes for chemistry. 
The irradiated gas reaches high temperatures (near 1000 K, \cite{champion2017herschel}) and a significant
fraction of the H$_2$ molecules are radiatively pumped through fluoresence to vibrationally excited states \cite{BlackJ_87a},
\mbox{H$_{2}^{*}$\,($v$$>$0)}.  This suprathermal excitation overcomes the endothermicity of reaction~[1], allowing H$_{2}^{*}$ to react with C$^+$, leading to the formation of abundant CH$^+$ \cite{Sternberg1995,Agundez2010, thi2011detection}. Subsequent fast and exothermic  hydrogen abstraction reactions 
\mbox{CH$^+$\,$\xrightarrow[(2)]{\rm H_2}$\,CH$_{2}^{+}$\,$\xrightarrow[(3)]{\rm H_2}$\,CH$_{3}^{+}$} then efficiently lead to CH$_{3}^{+}$.
{In the Methods section, we quantitatively assess these processes using models, and show that for a wide range of acceptable parameters, CH$_{3}^{+}$ is formed efficiently in FUV irradiated environments.}
The formed CH$_{3}^{+}$
reacts very slowly with H$_2$ (through radiative association) and is mainly destroyed by dissociative recombination  with electrons, leading to CH$_2$, CH, and C in comparable amounts \cite{Thomas2012}.
CH$_{3}^{+}$ can also be destroyed by reactions with neutral oxygen producing	
HCO$^+$ and with neutral molecules	producing molecular ions. These	undergo
dissociative recombination with	electrons yielding complex organic
molecules. Therefore, in the presence of UV radiation,  gas-phase
organic	chemistry is initiated through CH$_{3}^{+}$ \cite{smith1992ion, herbst2021unusual,Cuadrado2015}.

The ongoing chemistry in d203-506 described above differs greatly from what has been observed in disks that are not 
exposed to external UV irradiation where the freeze out of H$_2$O and CO$_2$ control the gas composition.
In such disks, high abundances of water, HCN, CH$_4$, C$_2$H$_2$, etc.\ are observed 
\cite{pontoppidan2010spitzer, grant2023minds}, species which are not detected 
in d203-506. 
{In the last decades, the formation of organic molecules in space has been considered to happen mostly at the surface of grains \cite{cuppen2017grain, semenov2010chemistry}. The detection of CH$_3^+$ indicates that alternative gas-phase routes are available to activate the organic chemistry, when UV radiation is present. }
Far from being anecdotal, external UV irradiation is expected to 
occur during the early life of most protoplanetary disks \cite{winter2022external}, {making UV-driven organic chemistry
common  for the chemical evolution of most protoplanetary disks and of the early the Solar System  \cite{bergin2023interstellar, naraoka2023soluble}.}
More generally, {this chemistry can be active} in any environment providing 
sufficiently high gas density and FUV irradiation ($n_{\rm H} \gtrsim 10^5$ cm$^{-3}$, $G_0 \gtrsim 10^4$). 
This can include, for instance, star-forming regions, the envelopes of planetary nebulae, the inner regions of 
disks around T-Tauri stars, and the interstellar medium of star-forming galaxies near and far.
While the CH$_3^+$ detection presented here is a promising achievement, there are still numerous unanswered questions surrounding the excitation, chemistry, and spectroscopic properties of this species (Methods and Supplementary Information). {These topics shall be addressed thanks to interdisciplinary scientific efforts that incorporate the expertise of astronomers, physicists, and spectroscopists (both laboratory and theory) in order to fully understand the role of CH$_3^+$ in organic chemistry in space.}

\bibliographystyle{nature.bst} 
\bibliography{references}

\begin{thebibliography}{10}
\expandafter\ifx\csname url\endcsname\relax
  \def\url#1{\burl{#1}}\fi
\expandafter\ifx\csname urlprefix\endcsname\relax\def\urlprefix{URL }\fi
\providecommand{\bibinfo}[2]{#2}
\providecommand{\eprint}[2][]{\url{#2}}
\providecommand{\doi}[1]{\url{https://doi.org/#1}}

\bibitem{black1977models}
\bibinfo{author}{Black, J.} \& \bibinfo{author}{Dalgarno, A.}
\newblock \bibinfo{title}{Models of interstellar clouds. i-the zeta ophiuchi
  cloud}.
\newblock \emph{\bibinfo{journal}{Astrophys. J. Supplement Series}}
  \textbf{\bibinfo{volume}{34}}, \bibinfo{pages}{405--423}
  (\bibinfo{year}{1977}).

\bibitem{smith1992ion}
\bibinfo{author}{Smith, D.}
\newblock \bibinfo{title}{The ion chemistry of interstellar clouds}.
\newblock \emph{\bibinfo{journal}{Chemical reviews}}
  \textbf{\bibinfo{volume}{92}}, \bibinfo{pages}{1473--1485}
  (\bibinfo{year}{1992}).

\bibitem{herbst2021unusual}
\bibinfo{author}{Herbst, E.}
\newblock \bibinfo{title}{Unusual chemical processes in interstellar chemistry:
  Past and present}.
\newblock \emph{\bibinfo{journal}{Frontiers in Astronomy and Space Sciences}}
  \textbf{\bibinfo{volume}{8}}, \bibinfo{pages}{776942} (\bibinfo{year}{2021}).

\bibitem{roueff2013ch2d+}
\bibinfo{author}{Roueff, E.} \emph{et~al.}
\newblock \bibinfo{title}{\ch{CH2D+}, the search for the holy grail}.
\newblock \emph{\bibinfo{journal}{The Journal of Physical Chemistry A}}
  \textbf{\bibinfo{volume}{117}}, \bibinfo{pages}{9959--9967}
  (\bibinfo{year}{2013}).

\bibitem{indriolo2010constraining}
\bibinfo{author}{Indriolo, N.}, \bibinfo{author}{Oka, T.},
  \bibinfo{author}{Geballe, T.} \& \bibinfo{author}{McCall, B.~J.}
\newblock \bibinfo{title}{Constraining the environment of \ch{CH+} formation
  with \ch{CH3+} observations}.
\newblock \emph{\bibinfo{journal}{Astrophys. J.}}
  \textbf{\bibinfo{volume}{711}}, \bibinfo{pages}{1338} (\bibinfo{year}{2010}).

\bibitem{cuppen2017grain}
\bibinfo{author}{Cuppen, H.} \emph{et~al.}
\newblock \bibinfo{title}{Grain surface models and data for astrochemistry}.
\newblock \emph{\bibinfo{journal}{Space Science Reviews}}
  \textbf{\bibinfo{volume}{212}}, \bibinfo{pages}{1--58}
  (\bibinfo{year}{2017}).

\bibitem{semenov2010chemistry}
\bibinfo{author}{Semenov, D.} \emph{et~al.}
\newblock \bibinfo{title}{{Chemistry in disks-IV. Benchmarking gas-grain
  chemical models with surface reactions}}.
\newblock \emph{\bibinfo{journal}{Astron. Astrophys.}}
  \textbf{\bibinfo{volume}{522}}, \bibinfo{pages}{A42} (\bibinfo{year}{2010}).

\bibitem{pdrs4all2022}
\bibinfo{author}{Bern{\'e}, O.} \emph{et~al.}
\newblock \bibinfo{title}{{PDRs4All: A JWST early release science program on
  radiative feedback from massive stars}}.
\newblock \emph{\bibinfo{journal}{Publications of the Astronomical Society of
  the Pacific}} \textbf{\bibinfo{volume}{134}}, \bibinfo{pages}{054301}
  (\bibinfo{year}{2022}).

\bibitem{bally2000disks}
\bibinfo{author}{Bally, J.}, \bibinfo{author}{O’Dell, C.} \&
  \bibinfo{author}{McCaughrean, M.~J.}
\newblock \bibinfo{title}{Disks, microjets, windblown bubbles, and outflows in
  the orion nebula}.
\newblock \emph{\bibinfo{journal}{Astron. J.}} \textbf{\bibinfo{volume}{119}},
  \bibinfo{pages}{2919} (\bibinfo{year}{2000}).

\bibitem{MentenK_07}
\bibinfo{author}{{Menten}, K.~M.}, \bibinfo{author}{{Reid}, M.~J.},
  \bibinfo{author}{{Forbrich}, J.} \& \bibinfo{author}{{Brunthaler}, A.}
\newblock \bibinfo{title}{{The distance to the Orion Nebula}}.
\newblock \emph{\bibinfo{journal}{Astron. Astrophys.}}
  \textbf{\bibinfo{volume}{474}}, \bibinfo{pages}{515--520}
  (\bibinfo{year}{2007}).

\bibitem{hillenbrand2000constraints}
\bibinfo{author}{Hillenbrand, L.~A.} \& \bibinfo{author}{Carpenter, J.~M.}
\newblock \bibinfo{title}{Constraints on the stellar/substellar mass function
  in the inner orion nebula cluster}.
\newblock \emph{\bibinfo{journal}{Astrophys. J.}}
  \textbf{\bibinfo{volume}{540}}, \bibinfo{pages}{236} (\bibinfo{year}{2000}).

\bibitem{CunhadeMiranda2010}
\bibinfo{author}{Cunha~de Miranda, B.~K.} \emph{et~al.}
\newblock \bibinfo{title}{Threshold photoelectron spectroscopy of the methyl
  radical isotopomers, \ch{CH3}, \ch{CH2D}, \ch{CHD2} and \ch{CD3}: Synergy
  between vuv synchrotron radiation experiments and explicitly correlated
  coupled cluster calculations}.
\newblock \emph{\bibinfo{journal}{The Journal of Physical Chemistry A}}
  \textbf{\bibinfo{volume}{114}}, \bibinfo{pages}{4818--4830}
  (\bibinfo{year}{2010}).

\bibitem{asvany2018spectroscopy}
\bibinfo{author}{Asvany, O.}, \bibinfo{author}{Thorwirth, S.},
  \bibinfo{author}{Redlich, B.} \& \bibinfo{author}{Schlemmer, S.}
\newblock \bibinfo{title}{Spectroscopy of the low-frequency vibrational modes
  of \ch{CH3+} isotopologues}.
\newblock \emph{\bibinfo{journal}{Journal of Molecular Spectroscopy}}
  \textbf{\bibinfo{volume}{347}}, \bibinfo{pages}{1--6} (\bibinfo{year}{2018}).

\bibitem{Kraemer1991:CH3+}
\bibinfo{author}{Kraemer, W.} \& \bibinfo{author}{Špirko, V.}
\newblock \bibinfo{title}{Potential energy function and rotation-vibration
  energy levels of \ch{CH3+}}.
\newblock \emph{\bibinfo{journal}{Journal of Molecular Spectroscopy}}
  \textbf{\bibinfo{volume}{149}}, \bibinfo{pages}{235--241}
  (\bibinfo{year}{1991}).

\bibitem{Keceli2009:CH3+}
\bibinfo{author}{Keçeli, M.}, \bibinfo{author}{Shiozaki, T.},
  \bibinfo{author}{Yagi, K.} \& \bibinfo{author}{Hirata, S.}
\newblock \bibinfo{title}{Anharmonic vibrational frequencies and
  vibrationally-averaged structures of key species in hydrocarbon combustion:
  \ch{HCO+}, \ch{HCO}, \ch{HNO}, \ch{HOO}, \ch{HOO-}, \ch{CH3+}, and \ch{CH3}}.
\newblock \emph{\bibinfo{journal}{Molecular Physics}}
  \textbf{\bibinfo{volume}{107}}, \bibinfo{pages}{1283--1301}
  (\bibinfo{year}{2009}).

\bibitem{Crofton1988:CH3+nu3}
\bibinfo{author}{Crofton, M.~W.}, \bibinfo{author}{Jagod, M.},
  \bibinfo{author}{Rehfuss, B.~D.}, \bibinfo{author}{Kreiner, W.~A.} \&
  \bibinfo{author}{Oka, T.}
\newblock \bibinfo{title}{{Infrared spectroscopy of carbo‐ions. III. $\nu_3$
  band of methyl cation \ch{CH3+}}}.
\newblock \emph{\bibinfo{journal}{The Journal of Chemical Physics}}
  \textbf{\bibinfo{volume}{88}}, \bibinfo{pages}{666--678}
  (\bibinfo{year}{1988}).

\bibitem{Hierl1997}
\bibinfo{author}{{Hierl}, P.~M.}, \bibinfo{author}{{Morris}, R.~A.} \&
  \bibinfo{author}{{Viggiano}, A.~A.}
\newblock \bibinfo{title}{{Rate coefficients for the endothermic reactions C+ +
  H2 -> CH+ + H }}.
\newblock \emph{\bibinfo{journal}{\jcp}} \textbf{\bibinfo{volume}{106}},
  \bibinfo{pages}{10145--10152} (\bibinfo{year}{1997}).

\bibitem{Zanchet2013}
\bibinfo{author}{{Zanchet}, A.} \emph{et~al.}
\newblock \bibinfo{title}{{H$_{2}$(v = 0,1) + C$^{+}$($^{2}$ P) - > H+CH$^{+}$
  State-to-state Rate Constants for Chemical Pumping Models in Astrophysical
  Media}}.
\newblock \emph{\bibinfo{journal}{Astrophys. J.}}
  \textbf{\bibinfo{volume}{766}}, \bibinfo{pages}{80} (\bibinfo{year}{2013}).

\bibitem{champion2017herschel}
\bibinfo{author}{Champion, J.} \emph{et~al.}
\newblock \bibinfo{title}{Herschel survey and modelling of
  externally-illuminated photoevaporating protoplanetary disks}.
\newblock \emph{\bibinfo{journal}{Astron. Astrophys.}}
  \textbf{\bibinfo{volume}{604}}, \bibinfo{pages}{A69} (\bibinfo{year}{2017}).

\bibitem{BlackJ_87a}
\bibinfo{author}{{Black}, J.~H.}, \bibinfo{author}{{Chaffee}, F.~H.} \&
  \bibinfo{author}{{Foltz}, C.~B.}
\newblock \bibinfo{title}{{Molecules at early epochs. II - H2 and CO toward PHL
  957}}.
\newblock \emph{\bibinfo{journal}{Astrophys. J.}}
  \textbf{\bibinfo{volume}{317}}, \bibinfo{pages}{442--449}
  (\bibinfo{year}{1987}).

\bibitem{Sternberg1995}
\bibinfo{author}{{Sternberg}, A.} \& \bibinfo{author}{{Dalgarno}, A.}
\newblock \bibinfo{title}{{Chemistry in Dense Photon-dominated Regions}}.
\newblock \emph{\bibinfo{journal}{Astrophys. J.s}}
  \textbf{\bibinfo{volume}{99}}, \bibinfo{pages}{565} (\bibinfo{year}{1995}).

\bibitem{Agundez2010}
\bibinfo{author}{{Ag{\'u}ndez}, M.}, \bibinfo{author}{{Goicoechea}, J.~R.},
  \bibinfo{author}{{Cernicharo}, J.}, \bibinfo{author}{{Faure}, A.} \&
  \bibinfo{author}{{Roueff}, E.}
\newblock \bibinfo{title}{{The Chemistry of Vibrationally Excited H$_{2}$ in
  the Interstellar Medium}}.
\newblock \emph{\bibinfo{journal}{Astrophys. J.}}
  \textbf{\bibinfo{volume}{713}}, \bibinfo{pages}{662--670}
  (\bibinfo{year}{2010}).

\bibitem{thi2011detection}
\bibinfo{author}{Thi, W.-F.} \emph{et~al.}
\newblock \bibinfo{title}{Detection of ch+ emission from the disc around hd
  100546}.
\newblock \emph{\bibinfo{journal}{Astron. Astrophys.}}
  \textbf{\bibinfo{volume}{530}}, \bibinfo{pages}{L2} (\bibinfo{year}{2011}).

\bibitem{Thomas2012}
\bibinfo{author}{{Thomas}, R.~D.} \emph{et~al.}
\newblock \bibinfo{title}{{Dissociative Recombination of Vibrationally Cold
  CH$^{+}$ $_{3}$ and Interstellar Implications}}.
\newblock \emph{\bibinfo{journal}{Astrophys. J.}}
  \textbf{\bibinfo{volume}{758}}, \bibinfo{pages}{55} (\bibinfo{year}{2012}).

\bibitem{Cuadrado2015}
\bibinfo{author}{{Cuadrado}, S.} \emph{et~al.}
\newblock \bibinfo{title}{{The chemistry and spatial distribution of small
  hydrocarbons in UV-irradiated molecular clouds: the Orion Bar PDR}}.
\newblock \emph{\bibinfo{journal}{Astron. Astrophys.}}
  \textbf{\bibinfo{volume}{575}}, \bibinfo{pages}{A82} (\bibinfo{year}{2015}).

\bibitem{pontoppidan2010spitzer}
\bibinfo{author}{Pontoppidan, K.~M.} \emph{et~al.}
\newblock \bibinfo{title}{A spitzer survey of mid-infrared molecular emission
  from protoplanetary disks. i. detection rates}.
\newblock \emph{\bibinfo{journal}{Astrophys. J.}}
  \textbf{\bibinfo{volume}{720}}, \bibinfo{pages}{887} (\bibinfo{year}{2010}).

\bibitem{grant2023minds}
\bibinfo{author}{Grant, S.~L.} \emph{et~al.}
\newblock \bibinfo{title}{Minds. the detection of 13co2 with jwst-miri
  indicates abundant co2 in a protoplanetary disk}.
\newblock \emph{\bibinfo{journal}{The Astrophysical Journal Letters}}
  \textbf{\bibinfo{volume}{947}}, \bibinfo{pages}{L6} (\bibinfo{year}{2023}).

\bibitem{winter2022external}
\bibinfo{author}{Winter, A.~J.} \& \bibinfo{author}{Haworth, T.~J.}
\newblock \bibinfo{title}{The external photoevaporation of planet-forming
  discs}.
\newblock \emph{\bibinfo{journal}{The European Physical Journal Plus}}
  \textbf{\bibinfo{volume}{137}}, \bibinfo{pages}{1132} (\bibinfo{year}{2022}).

\bibitem{bergin2023interstellar}
\bibinfo{author}{Bergin, E.~A.}, \bibinfo{author}{Alexander, C.},
  \bibinfo{author}{Drozdovskaya, M.}, \bibinfo{author}{Gounelle, M.} \&
  \bibinfo{author}{Pfalzner, S.}
\newblock \bibinfo{title}{Interstellar heritage and the birth environment of
  the solar system}  (\bibinfo{year}{To appear in ``Comets III'', University of
  Arizona Press}).

\bibitem{naraoka2023soluble}
\bibinfo{author}{Naraoka, H.} \emph{et~al.}
\newblock \bibinfo{title}{Soluble organic molecules in samples of the
  carbonaceous asteroid (162173) ryugu}.
\newblock \emph{\bibinfo{journal}{Science}} \textbf{\bibinfo{volume}{379}},
  \bibinfo{pages}{eabn9033} (\bibinfo{year}{2023}).

\bibitem{western2017:PGOPHER}
\bibinfo{author}{Western, C.~M.}
\newblock \bibinfo{title}{{{PGOPHER}}: {{A}} program for simulating rotational,
  vibrational and electronic spectra}.
\newblock \emph{\bibinfo{journal}{J. Quant. Spectrosc. Radiat. Transf.}}
  \textbf{\bibinfo{volume}{186}}, \bibinfo{pages}{221--242}
  (\bibinfo{year}{2017}).

\bibitem{PDRs4All_22}
\bibinfo{author}{{Bern{\'e}}, O.} \emph{et~al.}
\newblock \bibinfo{title}{{PDRs4All: A JWST Early Release Science Program on
  Radiative Feedback from Massive Stars}}.
\newblock \emph{\bibinfo{journal}{PASP}} \textbf{\bibinfo{volume}{134}},
  \bibinfo{pages}{054301} (\bibinfo{year}{2022}).

\bibitem{labiano2021wavelength}
\bibinfo{author}{Labiano, A.} \emph{et~al.}
\newblock \bibinfo{title}{Wavelength calibration and resolving power of the
  jwst miri medium resolution spectrometer}.
\newblock \emph{\bibinfo{journal}{Astron. Astrophys.}}
  \textbf{\bibinfo{volume}{656}}, \bibinfo{pages}{A57} (\bibinfo{year}{2021}).

\bibitem{BerneO_22}
\bibinfo{author}{{Bern{\'e}}, O.}, \bibinfo{author}{{Foschino}, S.},
  \bibinfo{author}{{Jalabert}, F.} \& \bibinfo{author}{{Joblin}, C.}
\newblock \bibinfo{title}{{Contribution of polycyclic aromatic hydrocarbon
  ionization to neutral gas heating in galaxies: model versus observations}}.
\newblock \emph{\bibinfo{journal}{Astron. Astrophys.}}
  \textbf{\bibinfo{volume}{667}}, \bibinfo{pages}{A159} (\bibinfo{year}{2022}).

\bibitem{roueffH2}
\bibinfo{author}{{Roueff}, E.} \emph{et~al.}
\newblock \bibinfo{title}{{The full infrared spectrum of molecular hydrogen}}.
\newblock \emph{\bibinfo{journal}{Astron. Astrophys.}}
  \textbf{\bibinfo{volume}{630}}, \bibinfo{pages}{A58} (\bibinfo{year}{2019}).

\bibitem{Foscino_19}
\bibinfo{author}{{Foschino}, S.}, \bibinfo{author}{{Bern{\'e}}, O.} \&
  \bibinfo{author}{{Joblin}, C.}
\newblock \bibinfo{title}{{Learning mid-IR emission spectra of polycyclic
  aromatic hydrocarbon populations from observations}}.
\newblock \emph{\bibinfo{journal}{Astron. Astrophys.}}
  \textbf{\bibinfo{volume}{632}}, \bibinfo{pages}{A84} (\bibinfo{year}{2019}).

\bibitem{TaboneB_21}
\bibinfo{author}{{Tabone}, B.}, \bibinfo{author}{{van Hemert}, M.~C.},
  \bibinfo{author}{{van Dishoeck}, E.~F.} \& \bibinfo{author}{{Black}, J.~H.}
\newblock \bibinfo{title}{{OH mid-infrared emission as a diagnostic of H$_{2}$O
  UV photodissociation. I. Model and application to the HH 211 shock}}.
\newblock \emph{\bibinfo{journal}{Astron. Astrophys.}}
  \textbf{\bibinfo{volume}{650}}, \bibinfo{pages}{A192} (\bibinfo{year}{2021}).

\bibitem{ZanneseM_22}
\bibinfo{author}{{Zannese}, M.} \emph{et~al.}
\newblock \bibinfo{title}{{OH mid-infrared emission as a diagnostic of H$_2$O
  UV photodissociation. II. Application to interstellar PDRs}}.
\newblock \emph{\bibinfo{journal}{arXiv e-prints}}
  \bibinfo{pages}{arXiv:2208.13619} (\bibinfo{year}{2022}).

\bibitem{pound2022photodissociation}
\bibinfo{author}{Pound, M.~W.} \& \bibinfo{author}{Wolfire, M.~G.}
\newblock \bibinfo{title}{The photodissociation region toolbox: Software and
  models for astrophysical analysis}.
\newblock \emph{\bibinfo{journal}{Astron. J.}} \textbf{\bibinfo{volume}{165}},
  \bibinfo{pages}{25} (\bibinfo{year}{2022}).

\bibitem{gordon2022:HITRAN2020}
\bibinfo{author}{Gordon, I.} \emph{et~al.}
\newblock \bibinfo{title}{The {{HITRAN2020}} molecular spectroscopic database}.
\newblock \emph{\bibinfo{journal}{Journal of Quantitative Spectroscopy and
  Radiative Transfer}} \textbf{\bibinfo{volume}{277}}, \bibinfo{pages}{107949}
  (\bibinfo{year}{2022}).

\bibitem{Schulenburg2006:photoionisation}
\bibinfo{author}{Schulenburg, A.~M.}, \bibinfo{author}{Alcaraz, C.},
  \bibinfo{author}{Grassi, G.} \& \bibinfo{author}{Merkt, F.}
\newblock \bibinfo{title}{Rovibrational photoionization dynamics of methyl and
  its isotopomers studied by high-resolution photoionization and photoelectron
  spectroscopy}.
\newblock \emph{\bibinfo{journal}{The Journal of Chemical Physics}}
  \textbf{\bibinfo{volume}{125}}, \bibinfo{pages}{104310}
  (\bibinfo{year}{2006}).

\bibitem{chai2008:Longrange}
\bibinfo{author}{Chai, J.-D.} \& \bibinfo{author}{{Head-Gordon}, M.}
\newblock \bibinfo{title}{Long-range corrected hybrid density functionals with
  damped atom\textendash atom dispersion corrections}.
\newblock \emph{\bibinfo{journal}{Phys. Chem. Chem. Phys.}}
  \textbf{\bibinfo{volume}{10}}, \bibinfo{pages}{6615} (\bibinfo{year}{2008}).

\bibitem{dunning1989:Gaussian}
\bibinfo{author}{Dunning, T.~H.}
\newblock \bibinfo{title}{Gaussian basis sets for use in correlated molecular
  calculations. {{I}}. {{The}} atoms boron through neon and hydrogen}.
\newblock \emph{\bibinfo{journal}{J. Chem. Phys.}}
  \textbf{\bibinfo{volume}{90}}, \bibinfo{pages}{1007--1023}
  (\bibinfo{year}{1989}).

\bibitem{woon1993:Gaussian}
\bibinfo{author}{Woon, D.~E.} \& \bibinfo{author}{Dunning, T.~H.}
\newblock \bibinfo{title}{Gaussian basis sets for use in correlated molecular
  calculations. {{III}}. {{The}} atoms aluminum through argon}.
\newblock \emph{\bibinfo{journal}{J. Chem. Phys.}}
  \textbf{\bibinfo{volume}{98}}, \bibinfo{pages}{1358--1371}
  (\bibinfo{year}{1993}).

\bibitem{Gaussian2016}
\bibinfo{author}{Frisch, M.~J.} \emph{et~al.}
\newblock \bibinfo{title}{Gaussian 16 {{Revision A}}.01}
  (\bibinfo{year}{2016}).

\bibitem{Pracna1993:linestrength}
\bibinfo{author}{Pracna, P.}, \bibinfo{author}{Spirko, V.} \&
  \bibinfo{author}{Kraemer, W.}
\newblock \bibinfo{title}{Ab initio study of linestrengths of
  vibration-rotation transitions of ammonia and methyl cations}.
\newblock \emph{\bibinfo{journal}{Journal of Molecular Spectroscopy}}
  \textbf{\bibinfo{volume}{158}}, \bibinfo{pages}{433--444}
  (\bibinfo{year}{1993}).

\bibitem{nyman2019infrared}
\bibinfo{author}{Nyman, G.} \& \bibinfo{author}{Yu, H.-G.}
\newblock \bibinfo{title}{Infrared vibrational spectra of \ch{CH3+} and its
  deuterated isotopologues}.
\newblock \emph{\bibinfo{journal}{AIP Advances}} \textbf{\bibinfo{volume}{9}},
  \bibinfo{pages}{095017} (\bibinfo{year}{2019}).

\bibitem{jagod1994infrared}
\bibinfo{author}{Jagod, M.-F.}, \bibinfo{author}{Gabrys, C.~M.},
  \bibinfo{author}{R{\"o}sslein, M.}, \bibinfo{author}{Uy, D.} \&
  \bibinfo{author}{Oka, T.}
\newblock \bibinfo{title}{Infrared spectrum of ch3+ involving high
  rovibrationai levels}.
\newblock \emph{\bibinfo{journal}{Canadian Journal of Physics}}
  \textbf{\bibinfo{volume}{72}}, \bibinfo{pages}{1192--1199}
  (\bibinfo{year}{1994}).

\bibitem{LePetitF_06}
\bibinfo{author}{{Le Petit}, F.}, \bibinfo{author}{{Nehm{\'e}}, C.},
  \bibinfo{author}{{Le Bourlot}, J.} \& \bibinfo{author}{{Roueff}, E.}
\newblock \bibinfo{title}{{A Model for Atomic and Molecular Interstellar Gas:
  The Meudon PDR Code}}.
\newblock \emph{\bibinfo{journal}{Astrophys. J. S.}}
  \textbf{\bibinfo{volume}{164}}, \bibinfo{pages}{506--529}
  (\bibinfo{year}{2006}).

\bibitem{Goicoechea2007}
\bibinfo{author}{{Goicoechea}, J.~R.} \& \bibinfo{author}{{Le Bourlot}, J.}
\newblock \bibinfo{title}{{The penetration of Far-UV radiation into molecular
  clouds}}.
\newblock \emph{\bibinfo{journal}{Astron. Astrophys.}}
  \textbf{\bibinfo{volume}{467}}, \bibinfo{pages}{1--14}
  (\bibinfo{year}{2007}).

\bibitem{Cardelli1989}
\bibinfo{author}{{Cardelli}, J.~A.}, \bibinfo{author}{{Clayton}, G.~C.} \&
  \bibinfo{author}{{Mathis}, J.~S.}
\newblock \bibinfo{title}{{The Relationship between Infrared, Optical, and
  Ultraviolet Extinction}}.
\newblock \emph{\bibinfo{journal}{Astrophys. J.}}
  \textbf{\bibinfo{volume}{345}}, \bibinfo{pages}{245} (\bibinfo{year}{1989}).

\bibitem{Birnstiel2018}
\bibinfo{author}{{Birnstiel}, T.} \emph{et~al.}
\newblock \bibinfo{title}{{The Disk Substructures at High Angular Resolution
  Project (DSHARP). V. Interpreting ALMA Maps of Protoplanetary Disks in Terms
  of a Dust Model}}.
\newblock \emph{\bibinfo{journal}{Astrophys. J. Let.}}
  \textbf{\bibinfo{volume}{869}}, \bibinfo{pages}{L45} (\bibinfo{year}{2018}).

\bibitem{Walsh2013}
\bibinfo{author}{{Walsh}, C.}, \bibinfo{author}{{Millar}, T.~J.} \&
  \bibinfo{author}{{Nomura}, H.}
\newblock \bibinfo{title}{{Molecular Line Emission from a Protoplanetary Disk
  Irradiated Externally by a Nearby Massive Star}}.
\newblock \emph{\bibinfo{journal}{Astrophys. J. Let.}}
  \textbf{\bibinfo{volume}{766}}, \bibinfo{pages}{L23} (\bibinfo{year}{2013}).

\bibitem{Plasil2011}
\bibinfo{author}{{Plasil}, R.} \emph{et~al.}
\newblock \bibinfo{title}{{Reactions of Cold Trapped CH$^{+}$ Ions with Slow H
  Atoms}}.
\newblock \emph{\bibinfo{journal}{Astrophys. J.}}
  \textbf{\bibinfo{volume}{737}}, \bibinfo{pages}{60} (\bibinfo{year}{2011}).

\bibitem{Blint1976}
\bibinfo{author}{{Blint}, R.~J.}, \bibinfo{author}{{Marshall}, R.~F.} \&
  \bibinfo{author}{{Watson}, W.~D.}
\newblock \bibinfo{title}{{Calculations of the lower electronic states of
  CH$_{3}$$^{+}$: a postulated intermediate in interstellar reactions.}}
\newblock \emph{\bibinfo{journal}{Astrophys. J.}}
  \textbf{\bibinfo{volume}{206}}, \bibinfo{pages}{627--631}
  (\bibinfo{year}{1976}).

\bibitem{McEwan1999}
\bibinfo{author}{{McEwan}, M.~J.} \emph{et~al.}
\newblock \bibinfo{title}{{New H and H$_{2}$ Reactions with Small Hydrocarbon
  Ions and Their Roles in Benzene Synthesis in Dense Interstellar Clouds}}.
\newblock \emph{\bibinfo{journal}{Astrophys. J.}}
  \textbf{\bibinfo{volume}{513}}, \bibinfo{pages}{287--293}
  (\bibinfo{year}{1999}).

\bibitem{adams1977reactions}
\bibinfo{author}{Adams, N.} \& \bibinfo{author}{Smith, D.}
\newblock \bibinfo{title}{Reactions of hydrocarbon ions with hydrogen and
  methane at 300 k}.
\newblock \emph{\bibinfo{journal}{Chemical Physics Letters}}
  \textbf{\bibinfo{volume}{47}}, \bibinfo{pages}{383--387}
  (\bibinfo{year}{1977}).

\bibitem{Larson1998}
\bibinfo{author}{{Larson}, {\r{A}}.} \emph{et~al.}
\newblock \bibinfo{title}{{Branching Fractions in Dissociative Recombination of
  CH$^{+}$$_{2}$}}.
\newblock \emph{\bibinfo{journal}{Astrophys. J.}}
  \textbf{\bibinfo{volume}{505}}, \bibinfo{pages}{459--465}
  (\bibinfo{year}{1998}).

\bibitem{Scott2000}
\bibinfo{author}{{Scott}, G. B.~I.}, \bibinfo{author}{{Milligan}, D.~B.},
  \bibinfo{author}{{Fairley}, D.~A.}, \bibinfo{author}{{Freeman}, C.~G.} \&
  \bibinfo{author}{{McEwan}, M.~J.}
\newblock \bibinfo{title}{{A selected ion flow tube study of the reactions of
  small C$_{m}$H$_{n}$$^{+}$ ions with O atoms}}.
\newblock \emph{\bibinfo{journal}{\jcp}} \textbf{\bibinfo{volume}{112}},
  \bibinfo{pages}{4959--4965} (\bibinfo{year}{2000}).

\bibitem{argyriou2023jwst}
\bibinfo{author}{Argyriou, I.} \emph{et~al.}
\newblock \bibinfo{title}{{JWST MIRI flight performance: The Medium-Resolution
  Spectrometer}}.
\newblock \emph{\bibinfo{journal}{arXiv preprint arXiv:2303.13469}}
  (\bibinfo{year}{2023}).

\end{thebibliography}

\newpage

\newpage

\section*{Acknowledgments}

OB is funded by a CNES APR program. MIRI data reduction is performed at the French MIRI centre of expertise with the support of CNES and the ANR-labcom INCLASS between IAS and the company ACRI-ST. Part of this work was supported by the Programme National “Physique et Chimie du Milieu Interstellaire” (PCMI) of CNRS/INSU with INC/INP co-funded by CEA and CNES.
Quantum chemical calculations were performed using HPC resources from the ``M\'esocentre'' computing center of CentraleSupélec and \'Ecole Normale Sup\'erieure Paris-Saclay supported by CNRS and R\'egion \^Ile-de-France (http://mesocentre.centralesupelec.fr/).
JRG and SC thank the Spanish MCINN for funding support under grant PID2019-106110GB-I00. JC and EP acknowledge support from the University of Western Ontario, the Institute for Earth and Space Exploration, the Canadian Space Agency, and the Natural Sciences and Engineering Research Council of Canada. The Cologne spectroscopy group acknowledges funding by the Deutsche Forschungsgemeinschaft DFG (CRC956, sub-project B2, ID 184018867) and the ERC AdG Missions (ID: 101020583). Work by YO and MR is
carried out within the Collaborative Research Centre 956, sub-project C1, funded by the 
DFG – project ID 184018867. CB is grateful for an appointment at NASA Ames Research Center through the San Jos\'e State University Research Foundation (80NSSC22M0107).  TO acknowledges support from JSPS Bilateral Program, Grant Number 120219939.

\section*{Author contributions}

O.B. found the signal in the data and led the analysis of the data and  write-up of the article. 
M.A.M.D., I.S., U.J., B.G., E.D., L.C., E.B., F.A., J.C., E.R., J.B., O.A., C.J., S.S., S.T., J.C., M.G., A.T., T.O., M.Z. conducted the spectroscopic analysis and participated in the write-up.  
M.A.M.D. created figures 3, S4, S5.   
I.S. created figures 1, 2, S1, S2, S3. 
I.S. and O.B. created figure S6. 
J.G. performed the chemical models and figures S7, S8 and participated in the write-up.
O.B., E.H., E.P. led the JWST observing program.
I.S., J.G., E.D., E.B., F.A., J.C., A.C., B.T., C.J., A.T., M.Z., A.A., J.B.S., C.B., E.B., R.C., S.C., D.D., M.E., A.F., K.D., L.I., O.K., B.K., O.L., D.L., R.L.G., A.M., R.M., Y.O., T.O., S.P., M.P., M.R., M.R., B.S., T.S., A.S., B.T., D.V.P., S.V. and M.W. contributed to the observing program with JWST. 
I.S., A.C., R.C., A.S., B.T., F.A., D.V.P. reduced the data.
E.D., M.A.M.D., L.C., J.G. and O.B. conducted the column density analysis.
J.H.B. wrote the section on the excitation of \ch{CH3+}.
M.W. and J.H.B. corrected the English throughout the manuscript.
All authors contributed to the discussions and provided feedback on the manuscript.

\section*{Methods}

\subsection*{Observations and data reduction}

The JWST/MIRI Medium Resolution Spectroscopy (MRS) Integral Field Unit (IFU) data were obtained on 2023 January 31 as part of the JWST ERS 1288 program
(PI: O. Berné, E. Habart, E. Peeters, \cite{PDRs4All_22}) referred to as ``PDRs4All''. 
All four channels and the three sub-channels were used, covering a wavelength range of 4.9--28 \um at a spectral resolution of 4000--1500 \cite{labiano2021wavelength}.
The observations are centered on RA=05$^{\rm h}$35$^{\rm m}$ 20$^{\rm s}$.4749 DEC$=-05^\circ\, 25'\,, 10''.45$ and span a mosaic of 9 pointings.
The overall science exposure time is 14086.11s for the whole mosaic.
We used the FASTR1 readout mode with 4-point dithering.
We reduced the data using the JWST Data Reduction pipeline version 1.9.5.
The stage 2 residual fringe correction was applied in addition to the standard fringe correction step. 
A master background subtraction was applied in stage 3 of the reduction.
At the end of the data reduction, we obtained four MIRI datacubes, one for each channel, each channel containing its three corresponding sub-channels (short, medium and long). 
{ The details of the data reduction for MIRI-IFU is part
of a dedicated paper by the PDRs4All team (Chown et al. in preparation).}

In this paper we also use one narrow band image of the NIRSpec observations of the same object as well as a NIRCam filter. 
Specifically, we use the NIRCam F212N filter image and the NIRSpec spectral cube corresponding to the F290LP filter which spans from $\sim2.9$ to 5~\um.
Observations were obtained with JWST-NIRSpec (JWST-NIRCam) on 2022 September 10 and reduced using the JWST pipeline version 1.9.4 (1.7.1) with Calibration Reference Data System (CRDS) context file jwst\_1014.pmap (jwst\_0969.pmap). 
{ For the NIRCam observations, no OFF emission was subtracted. For the NIRSpec observations, a dedicated OFF observation was subtracted. 
The details of the data reduction for these two instruments is discussed in a dedicated paper by the PDRs4All team on NIRCam (Habart et al. in prep) and NIRSpec (Peeters et al. in prep). The details on the observing strategy can be found in \cite{BerneO_22}.}

\subsection*{Data analysis}

In order to work with complete spectra spanning all the MIRI MRS wavelength range of 4.9--28~\um, we first stitch the spectra of the four channels to remove jumps between spectral orders. 
For each spectrum, we arbitrarily choose the one from Channel 2 long as a reference. 
Shorter and longer wavelengths (Channels 1, 3, and 4) are thus scaled accordingly.
{
We create (with numpy) an array of wavelengths spanning the full wavelength range, 
i.e. between 4.900 and 27.901 \um, with 
50 000 points and a constant step .}
Each channel spectrum is then interpolated onto this grid and jumps are removed by scaling each spectral order based on the average flux in the overlap region. The integrated intensity is conserved in this procedure, and uncertainties propagated.

The two extracted spectra shown in blue and orange in Fig.~\ref{fig:spectrum} are from the ``ON'' and ||OFF'' position, respectively. 
The ON (OFF) spectrum were extracted from an ellipse (circular aperture) centered on the position $\alpha=$ 5:35:20.357 , $\delta=$ $-$5:25:05.81 ($\alpha=$ 5:35:20.370, $\delta=$ $-$5:25:04.97), with dimension $l=0.52$'', $h=0.38$'' (of radius $r=0.365$’’) and a position angle PA=+33 (0) degrees (trigonometric) with respect to North. 
In order to have the emission of d203-506 we choose to subtract the nebula emission by evaluating ON$-$OFF. 
The ON and OFF spectra over the full MIRI-MRS range are shown in Extended Data Fig.~\ref{fig:spectrum_sup}, and the full subtracted spectrum is shown in Extended Data Fig.~\ref{fig:cont_sub}. 
In Extended Data Fig.~\ref{fig:cont_sub}, some lines are negative due to the over-subtraction of ionized emission lines which dominate in the nebula but are absent in d203-506. Some PAH bands are seen in negative in the ON$-$OFF spectrum; this is due to intrinsic variation of the PAH bands due to e.g.~size or ionization, and this cannot be interpreted as PAH absorption.

\subsection*{Line identification}

Using the ON-OFF spectrum described previously and the line list provided by the PDRs4All ERS team \cite{pdrs4all2022}, we identified the strong emission lines present in the data. 
The main emission lines are from \HI\ and H$_2$ and are listed in Table~\ref{table:h1bis} and~\ref{table:h2bis}, respectively.
{The H$_2$ lines intensities presented in the latter table
are measured using a Gaussian fitting to the observed lines, and the wavelength of the 
H$_2$ transition from \cite{roueffH2}. The approach to fit the observed lines is presented in \cite{Foscino_19}.}
Nebular emission lines from atomic ions are also identified and are shown in Figures~\ref{fig:spectrum} and~\ref{fig:cont_sub} with black vertical 
lines with their name in the attached box.
In addition, a number of OH lines are also identified between 9 and 11 \um
and are shown as green vertical lines on the same Figures. 
For this wavelength range we used OH wavelengths from \cite{TaboneB_21} and \cite{ZanneseM_22}. The study of OH emission in d203-506 will be the subject of a forthcoming paper (Zannese et al. in prep.).

From the H$_2$ lines listed in Table~\ref{table:h2bis} we derive an excitation 
diagram, using the H$_2$ Toolbox \cite{pound2022photodissociation} developed as part 
of the PDRs4All project science enabling products (see \url{https://pdrs4all.org/seps/} and).
{This is a tool for fitting temperature, column density, 
and ortho-to-para ratio in H$_2$ excitation diagrams. A one or two temperature model 
is assumed, and the fit finds the excitation temperatures and column densities,
and optionally ortho-to-para ratio. The source code is available at 
\url{https://dustem.astro.umd.edu/tools.html}.} \cite
The result of this analysis is shown in Extended Data Fig. ~\ref{fig:h2rot}. The derived excitation temperature is $T_{\rm ex} = 923 \pm 48.2 K$. 

\subsection*{Other candidate molecules}

There are no unassigned series of lines observed in the 5.2--6.2~\um range, corresponding to \ch{C=O} or \ch{C+N} vibrations (Fig.~\ref{fig:spectrum}) thus excluding most small species containing these chemical functions as carrier of the observed
7 \um features. 
At longer wavelengths (8--17~\um, see Extended Data Fig.~\ref{fig:cont_sub}), the spectrum is devoid of strong unassigned emission lines. Many hydrocarbon molecules, radicals, and ions (e.g., \ch{CH2}, \ch{CH2+}, ...) possess low frequency modes and would thus emit in that range. Instead, the lowest vibrational modes of \ch{CH3+}  lie at 7 \um. 
We thoroughly inspected the literature data on other hydrocarbons and known interstellar species (both neutral and charged) for possible matches and used local thermodynamic equilibrium (LTE) models{} to predict the emission from molecules in this spectral range. The tested molecules include---not exhaustively---H$_2$O (and isotopologues), H$_2$O$^+$, NH$_4^+$, C$_2$H$_2$, \ch{CH3}, HCN, SO$_2$, all hydrocarbons present in the HITRAN database \cite{gordon2022:HITRAN2020}, NH$_3$, CH$_3$OH, and CH$_3$CN.

\subsection*{Spectroscopy of the Methyl cation}

\ch{CH3+} is a planar molecule belonging to the $D_{3h}$ group of symmetry. It possesses four fundamental modes of vibration following an irreducible representation $\Gamma = 1 A_1' + 1 A_2'' + 2 E'$ (the two $E'$ modes are doubly degenerate). 
It is an oblate symmetric-top molecule for which rotational energy levels {of non-degenerate vibrational states are described with two quantum numbers, $J$, the total-angular-momentum quantum number excluding nuclear spin, and $K$, the projection of $\vec{N}$ (the total angular momentum excluding nuclear and electron spins) along the principal axis of symmetry; an additional $l$ quantum number accounts for Coriolis-coupling in degenerate vibrational states ($l=\pm 1$ in $v_4=1$).}
{The energy levels in a given vibronic state} can be calculated using the energy of the vibronic state, two rotational constants ($A=B$ and $C$), centrifugal distortion parameters ($D_J$, $D_{JK}$, $D_K$...), and, for states of $E'$ symmetry, additional Coriolis-coupling and $l$-doubling parameters ($\zeta$, $\eta_J$, $q$).
The molecule possesses three equivalent hydrogen atoms (fermions, $I_\mathrm{H} = 1/2$) resulting in spin-statistical weights  of (0, 0, 4, 4, 2, 2) for the levels of the states of symmetries ($A_1', A_1'', A_2', A_2'', E', E''$) \cite{Schulenburg2006:photoionisation}.
In the following, we used the PGOPHER software \cite{western2017:PGOPHER} to simulate the rovibrational spectrum of \ch{CH3+}. 

Since no experimental information is available on the $v_2=1$ ($A_2''$ symmetry) and $v_4=1$ ($E'$ symmetry) states of \ch{CH3+}, except the band positions \cite{asvany2018spectroscopy,CunhadeMiranda2010}, we rely on quantum chemical calculations to estimate the rotational constants in these excited states. We carried out geometry optimization and anharmonic frequency analysis at the $\omega$B97X-D/cc-pVQZ level of theory \cite{chai2008:Longrange,dunning1989:Gaussian, woon1993:Gaussian} using the Gaussian 16 suite of electronic structure programs \cite{Gaussian2016}. The main results from these calculations are reported in Table~\ref{tab:cts}. 
The table also contains the calculated constants reported in the literature \cite{Kraemer1991:CH3+, Keceli2009:CH3+}.
Besides the rotational constants, the quantum chemical calculations also give us insights about the transition moments of the $\nu_2$ {(out-of-plane bending, ``umbrella" motion)} and $\nu_4$ {(in-plane bending)} bands. Our calculations predict transition moments of {0.084 D and 0.064 D} for $\nu_2$ and $\nu_4$, respectively. Such low values have to be taken with caution as {experimental} values can differ significantly{, by 0.1 D or more}. Nevertheless, these values are consistent with the small values obtained by CCSD(T)/cc-pwCVTZ calculations performed in course of the work presented in Ref.~\cite{asvany2018spectroscopy} (0.06 D and 0.07 D, respectively), as well as those reported in Ref.~\cite{Pracna1993:linestrength} (0.049 D and 0.111 D) and Ref. \cite{nyman2019infrared} (0.10 D and 0.16 D).

To assess the reliability of the spectroscopic constants derived from the different quantum chemical calculations (both from this work and the literature), the calculated values in $v=0$ and $v_3=1$ were compared to the experimental values of the $\nu_3$ band.  Transitions within the $\nu_3$ band {(asymmetric stretching)} observed by Crofton et al. \cite{Crofton1988:CH3+nu3} were fitted in the present work using PGOPHER (so as to use the same model for $\nu_3$ and $\nu_2/\nu_4$) leading to the spectroscopic parameters reported in Table~\ref{tab:cts}. These derived parameters are in excellent agreement with those reported in Ref. \cite{Crofton1988:CH3+nu3}. They also are in very good agreement with the parameters obtained by the anharmonic calculations from this study. These results were used to scale the calculated constants in $v_2=1$ and $v_4=1$ according to the formula $B_{\mathrm{scaled}\,v_i}=B_{\mathrm{calc}\,v_i} \times B_{\mathrm{exp}\,v_3=1} / B_{\mathrm{calc}\,v_3=1}$ (and similarly for $C$). For the rotational constants calculated in Ref. \cite{Kraemer1991:CH3+}, since no calculated values in $v_3=1$ are reported, the scaling was made using values in $v=0$. The scaled values appear in blue in Table \ref{tab:cts}; they serve as a range of confidence for the spectral simulations.

Despite our best efforts, no definite spectroscopic analysis of the bands observed by JWST was achieved, i.e., we could not assign with confidence quantum numbers to the observed transitions. Instead, we performed several simulations by varying the rotational constants of $v_2=1$ and $v_4=1$ until qualitative agreement with the observational data was achieved. {Such qualitative agreement was assessed using the following criteria (by decreasing order of importance): i) Q-branch line position and intensity; ii) P- and R-branches spread and spectral line density; iii)  line positions and intensities in the P- and R-branches.}
In all cases, the Coriolis interaction constants were kept fixed to the calculated values of those in $v_3=1$, {and so were the transition moments of the $\nu_2$ and $\nu_4$ bands}.  
This lead us to produce four models, I to IV (see Table \ref{tab:cts} and Figures \ref{fig:modelI}-\ref{fig:modelIII}; model III being the model presented in the main article), that reasonably reproduce the astronomical data. It is worth noting, however, that the spectral density is greater on the spectrum of d203-506 than in our simulations, which could either reflect the presence of another species, or some discrepancies in the rotational constants used in models I to IV. 
Regarding the temperature, for all models, a rotational temperature of 400 K seems to adequately simulate the astronomical features (i.e., the spread of the P-, Q- and R-branches). {Higher temperatures lead to P- and R-branches that spread further than what is observed in the astronomical spectrum.}
While the simulations were performed under the assumption of thermal equilibrium, astronomical excitation conditions may differ significantly (see detailed discussion in the following section). This is particularly relevant to the relative intensities of the $\nu_2$ and $\nu_4$ bands, since the $v_2=1$ and $v_4=1$ vibrational states may be populated differently in d203-506. 

We have also investigated the presence of lines emanating from the $\nu_3$ band of CH$_3^+$ near 3~\um in the NIRSpec data of d203-506 presented in Bern\'e et al.(in prep.). The $\nu_3$ band has been observed at high resolution in the laboratory by Refs.~\cite{Crofton1988:CH3+nu3,jagod1994infrared}. Using the constants derived from these observations, we computed a synthetic spectrum of the $\nu_3$ band at 400~K and compare it to the NIRSpec background subtracted spectrum (using the same apertures as for MIRI, see Methods) in Extended Data Fig.~\ref{fig:NIRSpec}. The predicted lines are found to coincide with lines in the observations. 

In summary, high-resolution laboratory infrared spectra of CH$_3^+$ in the 7~\um region are needed to identify individual transitions in the astronomical data. First laboratory measurements at low temperatures will initiate this process, hopefully deciphering the strong Coriolis coupling between the degenerate $\nu_2$ and $\nu_4$ vibrations of CH$_3^+$. In order to support a quantitative analysis of the astronomical spectra, laboratory works also conducted at higher temperatures are required.

\subsection*{Chemistry of the Methyl cation}

We modeled the chemistry in the   {strongly FUV--irradiated photoevaporative wind and upper disk layers} of  d203-506 using the Meudon Photodissociation Region (PDR hereafter) code \cite{LePetitF_06}. 
The code solves the FUV radiative transfer in a medium of gas and dust 
\cite{Goicoechea2007}, as well as the steady-state heating, cooling, chemistry,
and H$_2$\,($v$,\,$J$) level populations as a function of depth into the neutral disk (in mag of visual extinction $A_V$). Bern\'e et al.(in prep.) previously used
this code to reproduce the H$_2$ line intensities detected by JWST/NIRSpec and obtained a radiation field intensity $G_0$\,$\simeq$\,4$\times$10$^4$ and a gas density \mbox{$n_{\rm H}$\,=\,$n$(H)\,+\,2$n$(H$_2$)} $\simeq$\,3.5$\times$10$^5$\,cm$^{-3}$
as the best-fit. We first adopt the same physical parameters
and use an extinction law suited to Orion molecular cloud \cite{Cardelli1989}, and consistent with dust grains ($R_V$\,=\,5.62 and \mbox{$N_{\rm H}$\,/\,$E(B-V)$\,=\,1.05$\times$10$^{22}$\,cm$^{-2}$}) bigger than in standard diffuse interstellar clouds ($R_V$\,=\,3.1). {This choice leads to a FUV dust extinction cross-section,
\mbox{$\sigma_{\rm 1000A}$(FUV)\,=\,1.1$\times$10$^{-21}$\,cm$^{-2}$/H}, which is in the upper range of
the cross-sections estimated by Bern\'e et al. (in prep.) for this object, and a factor of about two smaller than standard ISM grains. In addition, we run models adopting "bigger grains" (by a factor of about four),  leading to \mbox{$\sigma_{\rm 1000A}$(FUV)\,=\,7$\times$10$^{-22}$\,cm$^{-2}$/H}. This
smaller FUV cross-section is still compatible with the wind models presented by Bern\'e et al. (in prep.) and with the kind of dust grains
expected in the upper layers of protoplanetary disks \cite{Birnstiel2018}}.

Extended Data Fig.~\ref{fig:chemistry} shows the predicted density and temperature structure  of the 
wind and upper disk layers (upper panels) as well as the C$^+$, CH$^+$, CH$_{2}^{+}$,  CH$_{3}^{+}$, CO,
and HCO$^+$ abundance profiles (lower panels). {Figure~\ref{fig:chemistry}a refers to models using ``Orion grains'' and \mbox{$n_{\rm H }$\,=\,\,3.5$\times$10$^5$\,cm$^{-3}$} (Bern\'e et al. in prep.). The other
plots refer to models adopting ``bigger grains'' and densities of
3.5$\times$10$^5$\,cm$^{-3}$ (Extended Data Fig.~\ref{fig:chemistry}b), 3.5$\times$10$^6$\,cm$^{-3}$ (Extended Data Fig.~\ref{fig:chemistry}c), and 10$^7$\,cm$^{-3}$ (Extended Data Fig.~\ref{fig:chemistry}d). The latter
ones are more representative of the outer layers of a disk \cite{Walsh2013}.}

{In all these models,} reaction \mbox{C$^+$\,+\,H$_2$($v$,\,$J$)\,$\rightarrow$\,CH$^+$\,$+$\,H} [1] 
drives the  formation of CH$^+$ as soon as the H$_2$ abundance rises. In these conditions, high temperatures and presence of \mbox{FUV-pumped} vibrationally excited H$_{2}^{*}$,  \mbox{reaction~[1]} is
much faster than the slow radiative association reactions
\mbox{C$^+$\,+\,H$_2$\,$\rightarrow$\,CH$_{2}^{+}$\,$+$\,photon} and
\mbox{C$^+$\,+\,H\,$\rightarrow$\,CH$^+$\,$+$\,photon}. {These radiative
associations produce small amounts of CH$_{2}^{+}$ and CH$^+$ in cold gas ($T$\,$<$\,100\,K)}.
Our models include an \mbox{H$_2$($v$,\,$J$)} state-dependent treatment of 
\mbox{reaction~[1]} \cite{Zanchet2013}, {appropriate to the nonthermal populations
of H$_{2}^{*}$ in \mbox{FUV-irradiated environments}}. In particular, the CH$^+$ formation rate is computed by summing over all formation rates for each specific state of H$_2$. Once CH$^+$ is formed, fast and exoergic hydrogen abstraction reactions 
 \mbox{CH$^+$\,$\xrightarrow[]{\rm H_2}$\,CH$_{2}^{+}$\,$\xrightarrow[]{\rm H_2}$\,CH$_{3}^{+}$} lead to CH$_{3}^{+}$.
The \textit{efficiency} of this  chemical pipe to CH$_{3}^+$ depends on the abundance of H atoms in the gas (because they readily react with CH$^+$; \cite{Plasil2011})  and that of electrons (because they  destroy CH$_{2}^+$ and CH$_{3}^+$). Reactions with
atomic hydrogen dominate CH$^+$ destruction when 
the molecular gas fraction, defined as $f_{\rm H_2}$\,=\,2$n$(H$_2$)/$n_{\rm H}$, is $\lesssim$\,0.5 ($f_{\rm H_2}$\,=\,1 when all hydrogen is in molecular form).
{All models in \mbox{Extended Data Fig.~\ref{fig:chemistry}} predict} that the CH$_{3}^{+}$ abundance peaks
 close to the H\,/H$_2$ transition, at \mbox{$A_V$\,$\simeq$\,1~mag}, where \mbox{$T$\,$\simeq$\,1000-800\,K}. Extended Data Fig.~\ref{fig:chemical_network} summarizes the dominant chemical reactions at the CH$_{3}^{+}$ abundance peak. We note that the CH$^+$ and CH$_{3}^+$ abundance profiles
roughly follow the density profile of vibrationally excited H$_{2}^{*}$ (dotted black curve in the upper panel of Extended Data Fig.~\ref{fig:chemistry}). 
Hence, {irrespective of the exact gas density value}, the infrared H$_{2}$ lines detected with JWST in \mbox{d203-506} probe strongly FUV-irradiated hot gas, where CH$^+$ and CH$_{3}^{+}$ efficiently form.

{Our models predict \mbox{$N$(CH$_{3}^{+}$)/$N$(CH$^+$)} column density ratios of $\simeq$\,1--15 (increasing as $n_{\rm H}$ increases). We note
that the bulk column density of these species stem from \mbox{FUV-illuminated} gas
at \mbox{$A_V$\,$<$\,3\,mag}. Deeper inside, their abundances drop by orders of magnitude. Hence, both CH$^+$ and CH$_{3}^{+}$ are chemical
tracers of the most irradiated wind and upper disk layers.}

The local CH$^+$\,/\,CH$_{3}^{+}$ abundance ratio can be analytically estimated from the following
 network of gas-phase chemical reactions:
\begin{center}
 \mbox{C$^+$\,+\,H$_2$\,($v$,\,$J$)\,$\rightarrow$\,CH$^+$\,$+$\,H} \hspace{3cm}[1]\\
 \mbox{CH$^+$\,+\,H\,$\rightarrow$\,C$^+$\,$+$\,H$_2$} \hspace{3.7cm} [2a]\\
 \mbox{CH$^+$\,+\,H$_2$\,$\rightarrow$\,CH$_{2}^{+}$\,$+$\,H} \hspace{3.6cm} [2b]\\
 \mbox{CH$_{2}^{+}$\,+\,H$_2$\,$\rightarrow$\,CH$_{3}^{+}$\,$+$\,H} \hspace{3.5cm} [3]\\
 \mbox{CH$_{2}^{+}$\,+\,e$^-$\,$\rightarrow$\,products} \hspace{3.7cm} [4]\\

\mbox{CH$_{3}^{+}$\,+\,e$^-$\,$\rightarrow$\,products} \hspace{3.7cm} [5]\\
\end{center}

We note that adopting the photodissociation rate of \cite{Blint1976},
CH$_{3}^{+}$ photodissociation is expected
to be much slower than dissociative recombination with electrons even in strong UV fields. 
Therefore, in steady-state one
obtains:
\begin{equation}
\frac{x({\rm CH_{3}^{+}})}{x({\rm CH^+})}=  \frac{k_{2b}f_{H_2}^{2}}{2k_5\,x_e\,(f_{H_2}+2x_e\,k_4/k_3)}\simeq \frac{k_{2b}}{2k_5\,x_e}f_{{\rm H_2}}
\end{equation}
where $x_e$\,=\,$n$(e$^-$)/$n_{\rm H}$ is the electron abundance. 
In the last step we assumed that CH$_{2}^{+}$ destruction by reactions with H$_2$ are much faster than dissociative recombinations with electrons at the CH$_{3}^{+}$ abundance peak (as confirmed by the model). In our model we used the following reaction rate coefficients: $k_{2b}$\,=\,1.2$\times$10$^{-9}$\,cm$^3$\,s$^{-1}$ \cite{McEwan1999},
$k_{3}$\,=\,1.6$\times$10$^{-9}$\,cm$^3$\,s$^{-1}$ \cite{adams1977reactions},
$k_{4}$\,=\,6.40$\times$10$^{-7}$\,$(T/300)^{-0.60}$\,cm$^3$\,s$^{-1}$ \cite{Larson1998}, and 
$k_{5}$\,=\,6.97$\times$10$^{-7}$\,$(T/300)^{-0.61}$\,cm$^3$\,s$^{-1}$ \cite{Thomas2012}.
Using representative values for the wind and upper disk layers of \mbox{d203-506};
{\mbox{$T$\,=\,900\,K}} and \mbox{$x_e$\,$\simeq$\,$x$(C$^+$)\,$\simeq$1.4$\times$10$^{-4}$},
one obtains 
{\mbox{$x$({\rm CH$_{3}^{+}$})\,/\,$x$({\rm CH$^+$})\,$\simeq$\,12\,$f_{\rm H_2}$} from Eq.\,(1)}. {This analytical abundance ratio agrees with the
detailed predictions of our photochemical models. That is, the above chemical
reactions dominate the formation of CH$_{3}^{+}$ in FUV-irradiated gas.
In particular, models predict that the CH$_{3}^{+}$ abundance peaks
at gas molecular fractions of \mbox{$f_{\rm H_2}$\,$=$\,0.3--0.5}
(pink dotted curves in Figs.~\ref{fig:chemistry})}.

Reaction~[5] produces CH$_2$, CH, and C in similar amounts \cite{Thomas2012}. These are key reactive intermediate species that trigger the chemistry of  carbon species \cite{Cuadrado2015}.
In addition, reaction \mbox{O\,+\,CH$_{3}^{+}$\,$\rightarrow$\,HCO$^+$\,+\,H$_2$} \cite{Scott2000} is a dominant source of HCO$^+$, and thus of CO, in these irradiated hot gas layers ( \mbox{Extended Data Fig.~\ref{fig:chemistry}} also shows the predicted HCO$^+$ abundance profile {in the photoevaporative wind and upper disk layers)}. The morphology of the observed  \mbox{HCO$^+$\,$J$\,=\,4-3} line emission
{(first detected by \cite{champion2017herschel} and then mapped with ALMA
at high angular resolution by (Bern\'e et al. in prep.))} resembles
that of vibrationally excited H$_2$ and CH$_3$$^+$ observed with JWST (Bern\'e et al. in prep.). 
It will be difficult to explain the presence of extended HCO$^+$ emission in these  {strongly irradiated} gas layers without the FUV-driven chemistry described here and tested by the presence of CH$_{3}^{+}$. {We note that this hot HCO$^+$ linked
to the extended H$_{2}^*$ emission is different from the HCO$^+$ present in lower and denser layers of protoplanetary disks and formed by standard ion-molecule chemistry \cite{Walsh2013}}.

Specific 2D models, better adapted to the geometry of the upper disk layers and \mbox{FUV-irradiated} wind, will be needed to fully understand the 
{density structure} and  abundance  distribution of the observed molecular emission with JWST.

\newpage 

\section*{Data Availability}

The JWST data presented in this paper is publicly available through the MAST online archive (\url{http://mast.stsci.edu}) using the PID 1288. The MIRI spectra presented in Fig.~\ref{fig:spectrum} and Extended Data Figs.~\ref{fig:spectrum_sup},\ref{fig:cont_sub} are available in ASCII format at 
https://doi.org/10.5281/zenodo.7989669. The p-Gopher files to create the model spectra of \ch{CH3+} are available via : https://doi.org/10.5281/zenodo.7993330. 

 \section*{Code Availability}

 The JWST pipeline used to produce the final data produducts presented in this article is available at \url{https://github.com/spacetelescope/jwst}. 
 The MEUDON PDR code is publicly available at \url{https://ism.obspm.fr/pdr_download.html}

\newpage

\section*{Author information}

Supplementary Information is available for this paper.

Correspondence and requests for materials should be addressed to Olivier Bern\'e. 

The authors declare no financial or non-financial competing interest.

\newpage 
\section*{Supplementary information}

\subsection*{Supplementary methods}

\subsubsection*{$\nu_3$ emission of CH$_3^+$}

{
CH$_3^+$ has a band ($\nu_3$) near 3\um. This spectral range was observed as part
of the PDRs4All ERS program with the NIRSpec instrument. These data are subject of 
other papers in preparation, however we present here an analysis of these data 
for d203-506 in the context of the CH$_3^+$ detection. 
The NIRSpec spectrum in the spectral range of the $\nu_3$ band is shown in Fig.~\ref{fig:NIRSpec}. The shaded region represents the $\pm 3 \sigma$ error interval of the data. This error interval has been computed as the sum of the errors provided by the JWST pipeline and an empirical error. The empirical error is
three times the standard deviation of noise in the 3.102 to 3.126 spectral range of NIRSpec, where no lines are present.
In the same figure we show a best fit model of the OH lines which are present in this spectral range. This is an LTE model at a temperature of 800K. A detailed model of the OH emission will be presented in a forthcoming paper (Zannese et al. in prep). The objective here is not to obtain the best possible model, but rather to assess the contribution of OH in this spectral range. The H$_2$ lines present in the spectral range have been fitted individually with Gaussians using the wavelengths provided by \cite{roueffH2}. The width is set by the instrumental resolution. 
The CH$_3^+$ model used here is Model III., at a temperature of 400K.
All the models have been convolved to the spectral resolution of NIRSpec which is 
2100 in this range, and re-sampled on the NIRSpec wavelength grid. 
The total model is the sum of the OH, H$_2$ and CH$_3^+$ models. 
It can be seen in Fig.~\ref{fig:NIRSpec} that there is a good agreement between the total model and the NIRSpec data. We note that the intensity of the detectable CH$_3^+$ signal is comparable with the 3$\sigma$ noise of NIRSpec. Several strong lines of CH$_3^+$, in particular at 3.157 or 3.223 \um do appear clearly in the data. 
However, at this stage, the noise level in the data is too high to provide a definitive identification of specific lines of CH$_3^+$ with NIRSpec. We are currently working on obtaining a higher SNR spectrum by improving the data reduction, and by performing additional observations.}

\subsubsection*{Excitation of CH$_3^+$ infrared bands}

Hereafter we estimate the rates of the competing processes that might influence the excitation and line emission of  CH$_3^+$. 

\noindent
{\it COLLISIONAL EXCITATION}. A vibrationally excited state of the molecule can be excited by inelastic collisions with H or H$_2$ with an upward rate coefficient 
$$q_{\ell u} = q_0 \exp(-hc{\bar\nu}/kT_k) g_u/g_{\ell}\;\;\;{\rm cm}^3\;{\rm s}^{-1} $$
where $T_{\rm k}$ is the kinetic temperature of the gas, $q_0$ is the corresponding collisional quenching rate of the band $u\to \ell$, with degeneracies $g_u$ and $g_{\ell}$ of the upper and lower states, respectively. The wavenumber  of the transition is ${\bar\nu}$. 
{In the disk wind, 
the H$_2$ pure rotational lines, which have a critical density lower than that of the gas density, are thermalized, hence $T_{\rm rot} (H_{2}) = T_{\rm k} \approx 900$ K.} 
The quenching rate coefficient might be as large as $10^{-11}$ cm$^3$ s$^{-1}$; therefore, at the gas density of the reference PDR model shown in Fig.~\ref{fig:chemistry}, $n_{\rm H}=3.5\times 10^5$ cm$^{-3}$, the collisional excitation rates of the CH$_{3}^{+}$~~$\nu_4$ and $\nu_3$ fundamental bands would be at most $7\times 10^{-7}$ s$^{-1}$ and $5\times 10^{-8}$ s$^{-1}$ per ion, respectively. 
 {For a higher density of $n_{\rm H}=10^7$ cm$^{-3}$, these rates would be $2\times 10^{-5}$
and $1\times 10^{-6}$s$^{-1}$ per ion, respectively.}
The spontaneous transition probability of the $\nu_4$ fundamental band is calculated to be 
\mbox{$A=12$ s$^{-1}$}. This means that the vibrationally excited state of CH$_{3}^{+}$ cannot be thermally populated at the $T_{\rm k}$ : the collisional rates are many orders of magnitude too small, and thus $T_{\rm ex} < T_{\rm k}$.

\noindent
{\it FORMATION-PUMPING}.
As shown in previous sections, CH$_3^+$ is thought to form mainly through a sequence of exothermic ion-neutral reactions. The  change of enthalpy in the reaction
$$ {\rm CH}_2^+ + {\rm H}_2 \to {\rm CH}_3^+ +{\rm H} $$
is 65.6 kJ/mol, which corresponds to 5481 cm$^{-1}$. Energetically, this is sufficient to leave the product ion in the $v_4=1$ (1403 cm$^{-1}$) or $v_3=1$ (3108 cm$^{-1}$)  vibrationally excited states; however, this is likely to happen in only a small fraction $f_u$ of the reactions. In steady state, the total formation rate $F({\rm CH}_3^+)$ in cm$^{-3}$ s$^{-1}$ is balanced by a total destruction rate $n({\rm CH}_3^+) D$. The principal destruction processes in the PDR include dissociative recombination with electrons, slow radiative association reactions with H$_2$, reaction with atomic O to form HCO$^+$, and photodissociation. Photodissociation of CH$_3^+$ was investigated through {\it ab initio} computations \cite{Blint1976}, but there has been no recent treatment of the process. As a result, photodissociation of CH$_3^+$ is absent from the Leiden database of photodissociation and photoionization of astrophysically relevant molecules. \cite{Blint1976} identified the electronic states of CH$_3^+$ that would participate in its photodissociation and found only one  $^1$E'' state with an energy below 13.6 eV with the right symmetry. They estimated that the oscillator strength for transitions to this state must be small, $f\sim 10^{-3}$, implying that the photodissociation rate in the background Galactic radiation field must be of the order of 10$^{-11}$ s$^{-1}$. The reference PDR model in Fig.~\ref{fig:chemistry}  has $G_0=4\times 10^4$, so that that the corresponding photodissociation rate is $\sim 4\times 10^{-7}$ s$^{-1}$.  With reference to the zone of the PDR model where CH$_3^+$ is most abundant, $n_{\rm H}=3.5\times 10^5$ cm$^{-3}$, $T_{\rm k}=900$ K, and the electron fraction $x_e\approx 10^{-4}$. Adopting the rate coefficients of \cite{Thomas2012}
 for dissociative recombination, 
 we estimate that the destruction rate of CH$_3^+$ by electrons is $1.3\times 10^{-5}$ s$^{-1}$, which completely dominates over reactions with H$_2$ and O. {For higher densitites - as discussed in the previous section, this destruction process will be even more efficient. 
For a given density},  CH$_3^+$ is likely to be destroyed at least 100 times faster than it can be vibrationally excited by collisions at 900\,K. This means that the excited vibrational states cannot be populated in equilibrium. It also implies that the excitation rate by the formation process might be of the order of $\sim 10^{-5} f_u$ s$^{-1}$ per ion. Thus if the ion-neutral source reaction yields $\sim 1\%$ of the product CH$_3^+$ ions in $v_4=1$, the formation process could be at least as important as collisional excitation.

\noindent
{\it RADIATIVE EXCITATION}.
The rate of radiative pumping by absorption of continuous radiation is given by 
$$  {{A_{u\ell} g_u/g_{\ell}}\over{\exp(hc{\bar\nu}/kT_b)-1}} 
\;\;\;{\rm s}^{-1} $$
where $ A_{u\ell}$ is the spontaneous transition probability and $T_b$ is the brightness temperature of radiation at frequency $\bar\nu$ defined by the Planck function
$$ B_{\nu}(T_b) = {{2hc{\bar\nu}^3}\over{\exp(hc{\bar\nu}/kT_b)-1}} \;\;.$$   
As shown in Fig 2, the surface brightness of the continuum in the ON-OFF difference spectrum is 500 MJy sr$^{-1}$ ($=5\times 10^{-18}$ W m$^{-2}$ Hz$^{-1}$ sr$^{-1}$) near the wavelength of the $\nu_4$ fundamental band. This corresponds to $T_b=105$ K. In the calculation of Ref. \cite{nyman2019infrared}, this band has a transition dipole moment of $\mu=0.165$ Debye and $g_u/g_{\ell}=2$, so that the corresponding transition probability is $A=12$ s$^{-1}$. If the continuum source is roughly co-extensive with the CH$_3^+$-emitting region, then the molecules "feel" the same surface brightness that we see. Thus the radiative pumping rate in the $\nu_4$ band itself is of the order of $5\times 10^{-8}$ s$^{-1}$ per ion. The total continuum brightness in the PDR spectrum (Fig.~\ref{fig:spectrum}), is 10 times larger than that of the ON-OFF spectrum, which suggests that the molecules might be exposed to an even larger radiative excitation rate. The surface brightness in the NIRSpec spectrum of d$203-506$  is approximately 450 MJy sr$^{-1}$ at 3150 cm$^{-1}$ in the vicinity of the $\nu_3$ band, corresponding to $T_b=208$ K at this frequency. For a calculated transition probability $A=309$ s$^{-1}$ (dipole transition moment 0.181 Debye from \cite{nyman2019infrared}), this implies a radiative excitation rate of $1.1\times 10^{-7}$ s$^{-1}$ per ion. Both estimates of radiative excitation omit the diluted starlight from the stars that energize the PDR.

\noindent
{\it COMPARISON}. 
The rates of collisional excitation, hot-molecule formation, and infrared-radiative excitation of the observed CH$_{3}^{+}$ bands can be comparable to order-of-magnitude
in the case of a moderate density ($n_{\rm H}=3.5\times10^5$ cm$^{-3}$). 
{For higher densities ($n_{\rm H}=1\times10^7$ cm$^{-3}$) collisions are likely to dominate the excitation.} This example illustrates the possible non-LTE excitation mechanisms of the molecular vibrational bands that JWST may detect in related interstellar environments.

\subsubsection*{Column density estimates}

{Because the excitation processes that lead to the observed emission of CH$_3^+$ remain to be determined (see previous section), it is difficult at this stage to provide an accurate determination of the total column density of the species. We however provide an order of magnitude estimate hereafter, and also point to the main limiting factors in this derivation to guide future studies}. 

The integration over frequency of the emission band in the JWST observations in the 1200--1600 cm$^{-1}$ range, after local baseline subtraction {(spline fit on points outside the strong bands)}, is of $J_{\rm tot}\sim 8 \times 10^{-6}$ W/m$^2$/sr. To get an estimate of the column density of \ch{CH3+}, we operate under the assumption that the $\nu_2$ and $\nu_4$ bands of this ion are the major contributors to the observed features (once the strong isolated atomic and \ch{H2} lines are removed). Using the Einstein coefficients reported in Ref. \cite{Thomas2012} of $A(\nu_2) = 2.65$ s$^{-1}$ and $A(\nu_4) =$ 4.43 s$^{-1}$, {the column density in these upper emitting states can be derived by:}

\[  N_{\nu_2,\nu_4}\approx J_{\rm tot}\frac{4\pi}{h{\nu}\sum_{\nu_2,\nu_4}A} \] i.e., $ N_{\nu_2,\nu_4}\sim4.7\times10^{10}$~cm$^{-2}$. 
From PGOPHER models using { dipole moments values 0.084 and 0.064 D from our best model (III.)} for the vibrational modes $\nu_2$ and $\nu_4$, respectively, we derive a total emission in the 7 \um band of {$\rm 1.13\times10^{-19}$ W/CH$_3^+$}.  
Thus, under such assumptions a column density of {$N_{\nu_2,\nu_4}\sim7.8\times10^{10}$~cm$^{-2}$} is estimated. These numbers must be taken with caution however, {and are provided to the reader as an order of magnitude on the column densities for this species in this protoplanetary disk}, as the line strength are not determined experimentally yet, resulting in possible change by large factors in the derivation. {In addition,} as the excitation is unlikely to be at LTE {(see previous section)}, the determination of $N{_{\rm tot}}$(\ch{CH3+}), the total column density of \ch{CH3+}, cannot be safely determined by using simple Boltzmann factors to relate $\nu_2$ and $\nu_4$ populations to the ground state one. Within LTE approximation, this factor is given by the ratio of states degeneracies multiplied by $\exp(-h(\nu_i\rightarrow0)/kT)$. For the considered transitions, and excitation temperatures in the $800-300$~K range, it implies total column densities a factor of ten to thousand times {the above quoted} $N_{\nu_2,\nu_4}$, respectively. At 400~K at LTE, $N{_{\rm tot}}$(\ch{CH3+})$\, \sim 4\times10^{12}$~cm$^{-2}$.
A model including lines spectroscopic assignment and collisional rates, when they will be determined, is to be developed in the future to better constrain the \ch{CH3+} excitation (see section above) and the
 total column densities, under the observed astronomical conditions.
{Finally, an additional uncertainty in this derivation stems from the beam dilution, as the emission from the \ch{CH3+} emitting region is likely  not spatially resolved with MIRI. Hence -- put aside the spectroscopic uncertainties describe hereabove, the determined column density value should be considered as a lower limit.}
{Overall, this discussion further demonstrates the absolute need for collaborative actions coordinating astrophysicists observations and spectroscopists dedicated laboratory astrophysics experiments to interpret such sets of data.}

\newpage
\section*{Extended Data}

\setcounter{figure}{0}
\setcounter{table}{0}
\renewcommand{\thefigure}{\text{\arabic{figure}}}
\renewcommand{\thetable}{\text{\arabic{table}}}

\pagestyle{empty}

 \begin{figure*}
   \centering
   \includegraphics[width=16cm]{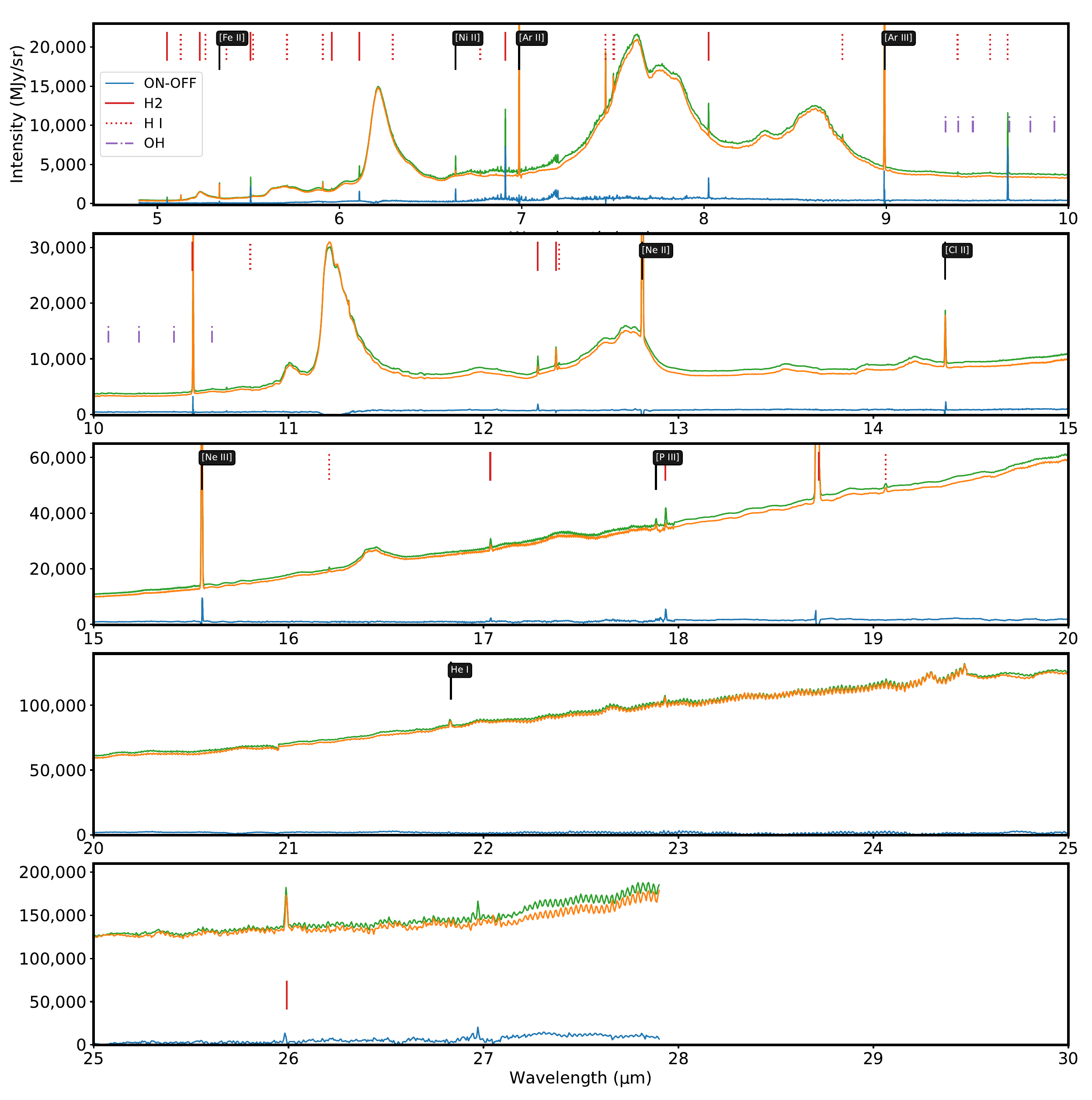}
      \caption{On and off spectra of d203-506 over the full MIRI-MRS spectral range. The On-Off spectrum is also shown. Main atomic and H$_2$ lines are indicated. 
              }
         \label{fig:spectrum_sup}
   \end{figure*}

\begin{figure*}
   \centering
   \includegraphics[width=16cm]{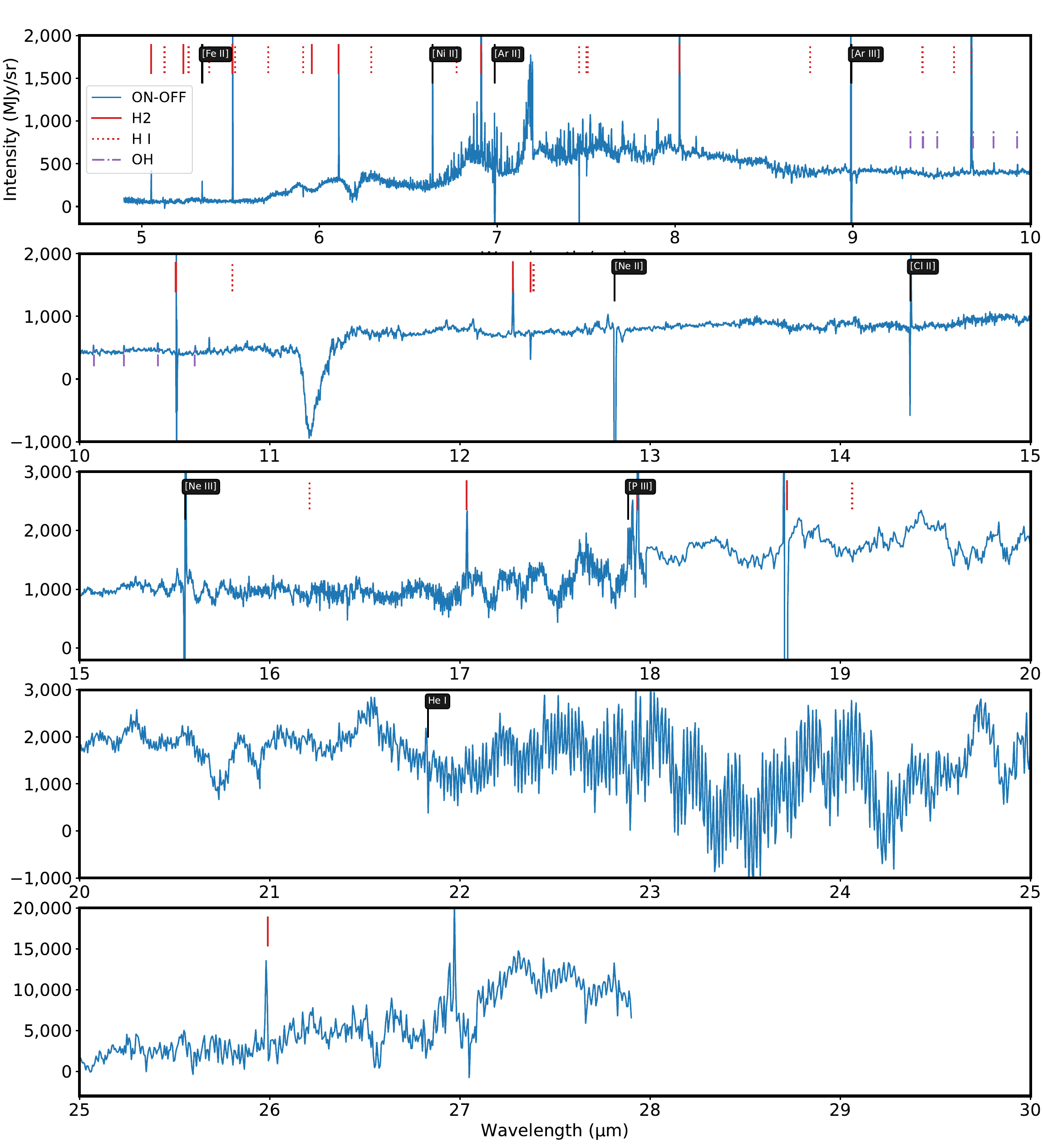}
      \caption{On-Off spectrum of d203-506 over the full MIRI-MRS spectral range. Main atomic and H$_2$ lines are indicated. 
              }
         \label{fig:cont_sub}
   \end{figure*}

 \begin{figure*}
   \centering
   \includegraphics[width=16cm]{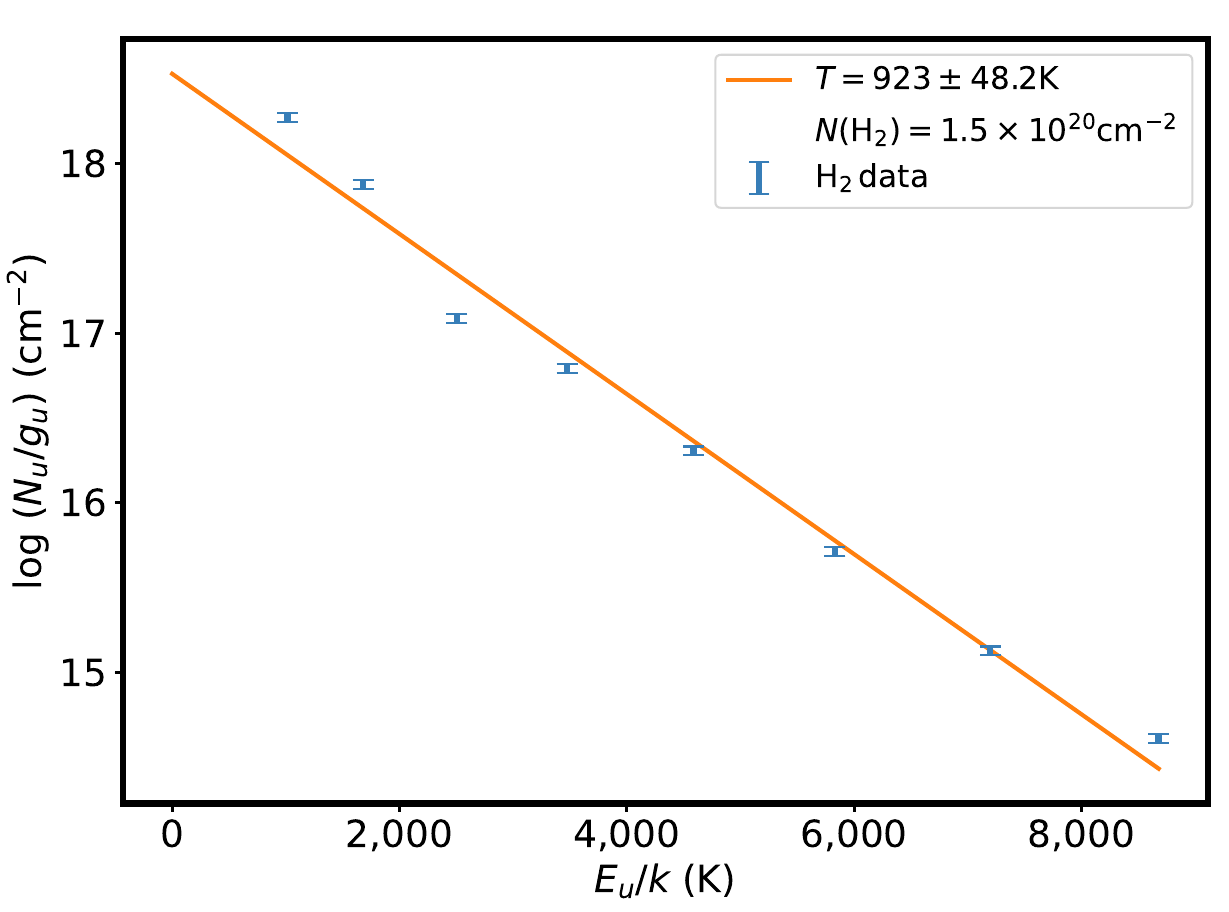}
   \caption{ H$_2$ excitation diagram derived from the line intensities in Table~\ref{table:h2bis} using the H$_2$ toolbox \cite{pound2022photodissociation} developed within the PDRs4All team as one of their Science Enabling Products. Error bars result from the propagation of the absolute calibration error of MIRI, which we take from \cite{argyriou2023jwst}.}
         \label{fig:h2rot}
   \end{figure*}

\FloatBarrier

\begin{table}[ht!]
\centering
\caption{\HI\ detected emission lines}
\label{table:h1bis}
\begin{tabular}{lccc}
\hline
$\lambda$ (\um) & $E_{\rm up}$ & $A$  &  Transition \\
(1) & (2) & (3)& (4) \\
\hline
5.128662 & 156225.1 & 3.6881 10$^4$&  6--10 \\
5.263685 & 157316.1 & 1.3121  10$^3$ & 7--18 \\
5.379776 & 157257.1 & 1.7812  10$^3$ & 7--17 \\
5.525190 & 157186.7 & 2.4709 10$^3$ &  7--16 \\
5.711464 & 157101.8 & 3.51558 10$^3$ &  7--15 \\
5.908220 & 155854.9 & 7.0652 10$^4$ & 6--9 \\
6.291918 & 156869.4 &  7.8457 10$^3$ &  7--13 \\
6.771993 & 156707.3 & 1.2503 10$^4$ & 7--12 \\
7.45984 & 153419.7 & 1.0254 10$^6$ & 5--6 \\
7.502502 & 155337.5 & 1.5609 10$^5$ & 6--8 \\
7.508107 & 156498.9 & 2.1174 10$^4$ & 7--11 \\
8.760068 & 156225.1 & 3.9049 10$^4$ & 7--10 \\
9.392013 & 156869.4 & 7.8037 10$^3$ & 8--13 \\
10.803593 & 157186.7 & 2.2679 10$^3$ & 9--16 \\
12.387158 & 156498.9 & 2.3007 10$^4$ & 8--11 \\
16.20909 & 156225.1 & 4.6762  10$^4$ & 8--10 \\
19.06192 & 155337.5 & 2.272 10$^5$ & 7--8 \\
\hline
\end{tabular}\\
{
(1) Emission line wavelength (\um);
(2) upper level energy (K);  
(3) Einstein A coefficient; (s$^{-1}$)
(4) Transition label.
}
\end{table}

\begin{table}[ht!]
\centering
\caption{Pure rotational H$_2$ detected emission lines in MIRI MRS wavelength range}
\label{table:h2bis}
\begin{tabular}{lccccr}
\hline
$\lambda$ (\um) & $T_u$ & $\nu$ & $A$ & Transition & Intensity ($\times 10^{-4}$) \\
(1) & (2) & (3) & (4) & (5) & (6)\\
\hline
5.0531 & 8677.1 & 1978.977 & 3.236 10$^{-7}$  & 0-0 S(8)   & 0.864 \\
5.5111 & 7196.7 & 1814.492 & 2.001 10$^{-7}$  & 0-0 S(7)   & 4.367  \\
6.1085 & 5829.8 & 1637.046 & 1.142 10$^{-7}$  & 0-0 S(6)   & 2.582  \\
6.9095 & 4586.1 & 1447.280 & 5.879 10$^{-8}$  & 0-0 S(5)   & 12.217  \\
8.0250 & 3474.5 & 1246.099 & 2.643 10$^{-8}$  & 0-0 S(4)   & 4.193  \\
9.6649 & 2503.7 & 1034.670 & 9.836 10$^{-9}$  & 0-0 S(3)   & 6.468  \\
12.278 & 1681.6 & 814.424  & 2.755 10$^{-9}$  & 0-0 S(2)   & 2.408  \\
17.034 & 1015.1 & 587.032  & 4.761 10$^{-10}$ & 0-0 S(1)   & 1.742  \\
\hline
\end{tabular}\\
{
(1) Emission line wavelength (\um), from \cite{roueffH2};
(2) Upper level energy (K);
(3) Transition energy (cm$^{-1}$);
(4) Einstein A coefficient (s$^{-1}$);
(5) transition label;
(6) erg~s$^{-1}$~cm$^{-2}$~sr$^{-1}$

}
\end{table}

\FloatBarrier

\begin{table}[p!]
\caption{Spectroscopic parameters of \ch{CH3+}.}
\label{tab:cts}
\scriptsize
\begin{tabular}{ll d{2.9} d{2.4}d{2.5}d{2.5} c d{2.6}d{2.6}d{2.6}d{2.6}}
\toprule
 &&       \multicolumn{1}{c}{\textbf{Experimental}}   & \multicolumn{3}{c}{\textbf{Calculated$^{*,a}$}}   && \multicolumn{4}{c}{\textbf{Models$^*$}}    \\   \cmidrule{4-6} \cmidrule{8-11}
 &&                                                   & \multicolumn{1}{c}{This work$^b$} & \multicolumn{1}{c}{Ref. \cite{Kraemer1991:CH3+}} & \multicolumn{1}{c}{Ref. \cite{Keceli2009:CH3+}} && \multicolumn{1}{c}{I}    & \multicolumn{1}{c}{II}    & \multicolumn{1}{c}{III} & \multicolumn{1}{c}{IV}   \\
\midrule
\multicolumn{3}{l}{$v=0$}   \\
&$B$      & 9.36214(28)    & 9.32        & 9.415        & 9.18        && 9.36214   & 9.36214   & 9.36214   & 9.36214  \\
&$C$      & 4.589949(35)    & 4.59        & 4.715        & 4.59        && 4.589949   & 4.589949   & 4.589949  & 4.589949  \\
&$D_{J}$  & 0.0007380(36)  & 0.00071     & 0.000719     &             && 0.0007380 & 0.0007380 & 0.0007380 & 0.0007380\\
&$D_{JK}$ & -0.0013144(79)  & -0.00124    & -0.001239    &             && -0.0013144 & -0.0013144 & -0.0013144 & -0.0013144\\
&$D_{K}$  & 0.0004552(51)  & 0.00057     & 0.000568     &             && 0.0004552  & 0.0004552  & 0.0004552  & 0.0004552 \\ \midrule
          
\multicolumn{3}{l}{$v_3=1$} \\
&$\nu$    & 3108.3556(18)  & \multicolumn{1}{c}{$2948$} &&&           &  \\
& $B$      & 9.27239(25)    & 9.21        &              & 9.00        &  \\
& $C$      & 4.550184(29)    & 4.46        &              & 4.50        &  \\
& $D_{J}$  & 0.0007029(30)  &             &              &             &  \\
& $D_{JK}$ & -0.0012814(71)  &             &              &             &  \\
& $D_{K}$  & 0.0004547(47)   &             &              &             &  \\
& $\zeta$  & 0.110551(38)    & 0.115       &              &             &  \\
& $\eta_J$ & -0.0006660(80)  &             &              &             &  \\
& $q^+$    & 0.00971(17)    &             &              &             &  \\  \midrule         
          
\multicolumn{3}{l}{$v_2=1$} \\
&$\nu$    & \multicolumn{1}{c}{$1372-1412$} & \multicolumn{1}{c}{$1412$} & \multicolumn{1}{c}{$1391,1433$} & \multicolumn{1}{c}{$1383,1418$} && 1392.80   & 1389.01  & 1391.22  & 1388.71 \\    
&$B$      &               & 9.21\,[\textcolor{blue}{9.27}]& 9.112\,[\textcolor{blue}{9.06}]      & 9.21\,[\textcolor{blue}{9.49}]             && 9.2270    & 9.3766   & 9.3721   & 9.3647  \\
&$C$      &               & 4.53\,[\textcolor{blue}{4.61}]& 4.758\,[\textcolor{blue}{4.63}]      & 4.61\,[\textcolor{blue}{4.66}]             && 4.6392    & 4.6542   & 4.6560   & 4.6651  \\
&$D_{J}$  &               &                               & 0.000715                            &                                            && 0.002201  & 0.000798 & .000703  & .000703 \\
&$D_{JK}$ &               &                               & -0.001212                           &                                            && -0.005267 & -0.00131 & -.00118  & -.00113 \\
&$D_{K}$  &               &                               & 0.000547                            &                                            && 0.002995  & 0.000488 & .000455  & .000455 \\ \midrule
         
\multicolumn{3}{l}{$v_4=1$} \\
&$\nu$    & \multicolumn{1}{c}{$1373-1393$} & \multicolumn{1}{c}{$1331$} & \multicolumn{1}{c}{$1399$} & \multicolumn{1}{c}{$1385,1429$}      && 1374.56   & 1374.46  & 1374.54  & 1396.35 \\
& $B$      &               & 9.44\,[\textcolor{blue}{9.50}]& 9.574\,[\textcolor{blue}{9.52}]     & 9.20\,[\textcolor{blue}{9.48}]             && 9.50      & 9.50     & 9.5000   & 9.5027  \\
& $C$      &               & 4.46\,[\textcolor{blue}{4.55}]& 4.743\,[\textcolor{blue}{4.62}]     & 4.60\,[\textcolor{blue}{4.65}]             && 4.5776    & 4.5802   & 4.5714   & 4.7534  \\
& $D_{J}$  &               &                               & 0.000719                            &                                            && 0.000747  & 0.000938 & .000703  & .000703 \\
& $D_{JK}$ &               &                               & -0.001240                           &                                            && -0.00116  & -0.00146 & -.00128  & -.00082 \\
& $D_{K}$  &               &                               & 0.000569                            &                                            && 0.000326  & 0.00032  & .000455  & .000455 \\ 
& $\zeta$  &               & 0.115                         & 0.1136                                &                                            && 0.11      & 0.11     & 0.11     & 0.11    \\
& $\eta_J$ &               &                               &                                     &                                            && -0.00063  & -0.00063 & -0.00063 & -0.00063\\
& $q^+$    &               &                               &                                     &                                            && 0.0095    & 0.0095   & 0.0095   & 0.0095  \\ \midrule
\multicolumn{3}{l}{$<v_2=1|v_4=1>^c$} & \\  
& $\zeta_{24}$  &               & -0.66                         &                                     &                                            && -0.66     & -0.66    & -0.66    & -0.66   \\        
\midrule
\multicolumn{3}{l}{$<v_i=1|d_i|v=0>^d$} & \\  
&$d_2$      &&& 0.084 & & & 0.084 & 0.084 & 0.084  & 0.084 \\
&$d_3$      &&& 0.102 & &  \\
&$d_4$      &&& 0.064 & & & 0.064 & 0.064 & 0.064 & 0.064 \\
\bottomrule
\end{tabular}

\medskip
\begin{minipage}{\textwidth}
$^*$ Rotational, centrifugal distortion, and Coriolis constants in the ground state, $v_3=1$, $v_2=1$, and $v_4=1$ (in cm$^{-1}$), and transition moments of the $\nu_2$ and $\nu_4$ bands. Comparison of quantum calculated  values, and those used to best model the spectrum of the d203-506 source around 1400 cm$^{-1}$.
$^a$ Values in brakets (in blue) are scaled according to $v_3=1$ results when available, to $v=0$ otherwise\\
$^b$ $\omega$B97X-D/cc-pVQZ, this work\\
$^c$ Coriolis interaction parameter between $v_2=1$ and $v_4=1$, unitless\\
$^d$ Transition dipole moment of the $\nu_i$ fundamental bands, in Debye
\end{minipage}
\end{table}

\FloatBarrier

\begin{figure}[ht!]
    \centering
    \includegraphics[width=\textwidth]{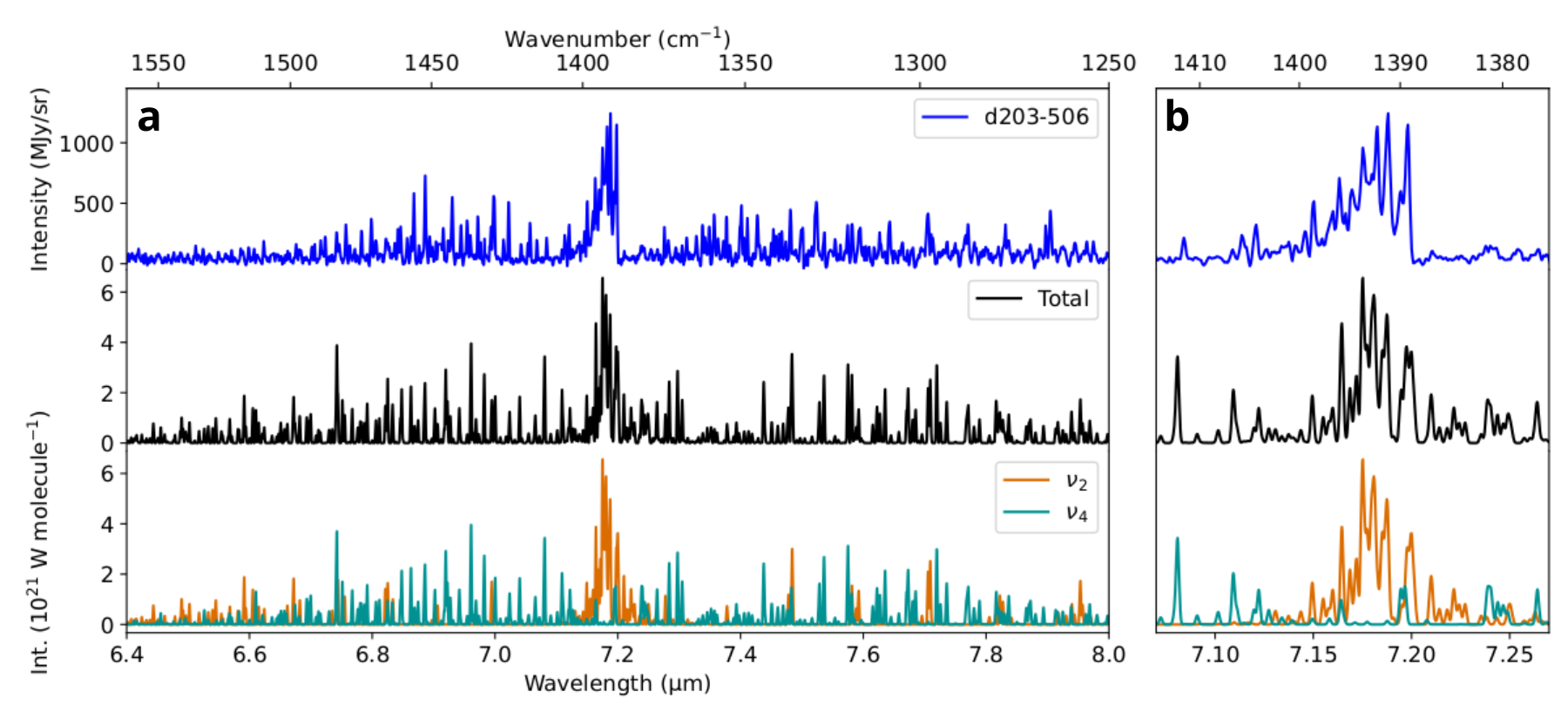}
    \includegraphics[width=\textwidth]{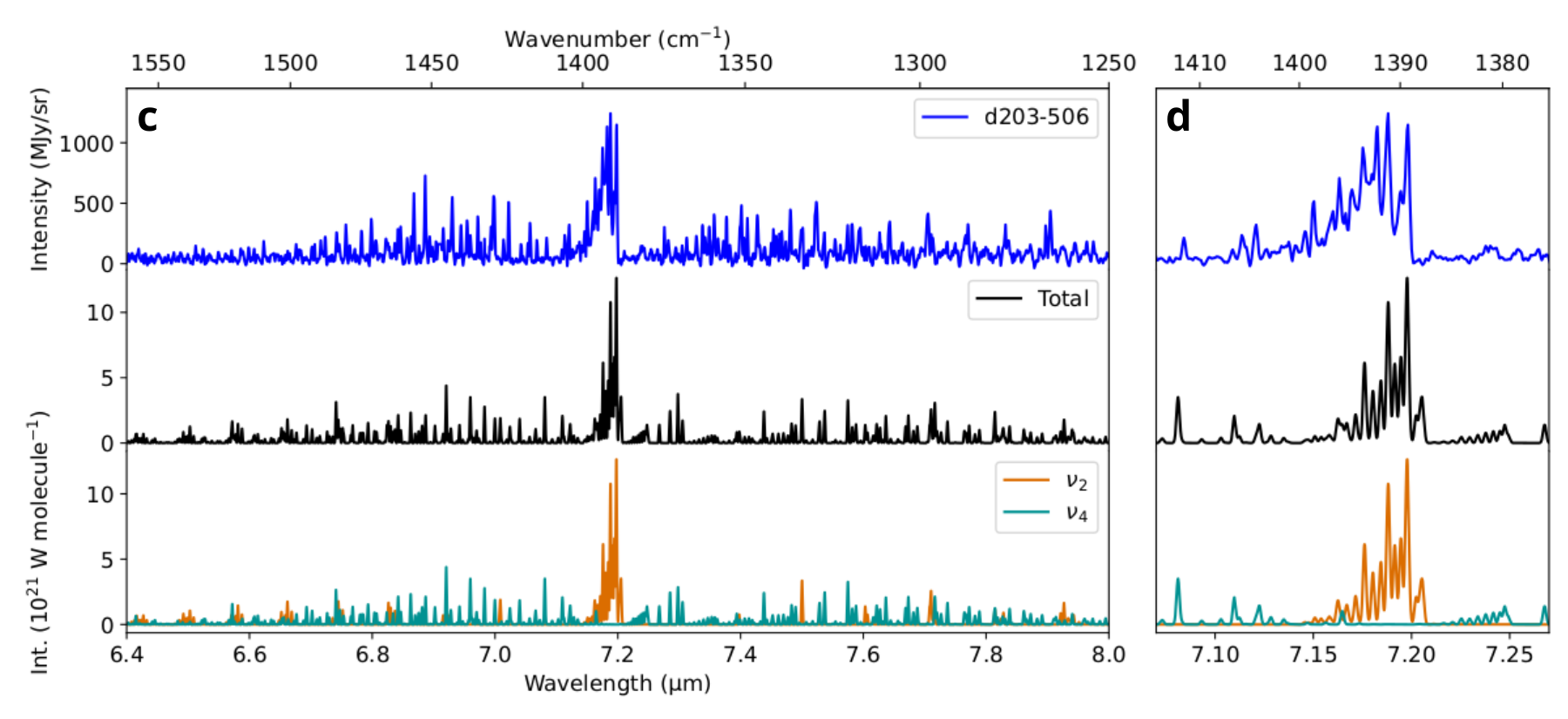}
    \caption{ {\bf a,} Model I, with zoom on the strongest lines ({\bf b}). {\bf c,} Model II, with zoom on strongest the lines ({\bf d}). For these models, the excitation temperature is $T=400$ K, and we use a Gaussian profile of 0.35 cm$^{-1}$ full-width-at-half-maximum.}
    \label{fig:modelI}
\end{figure}

\begin{figure}[ht!]
    \centering
    \includegraphics[width=\textwidth]{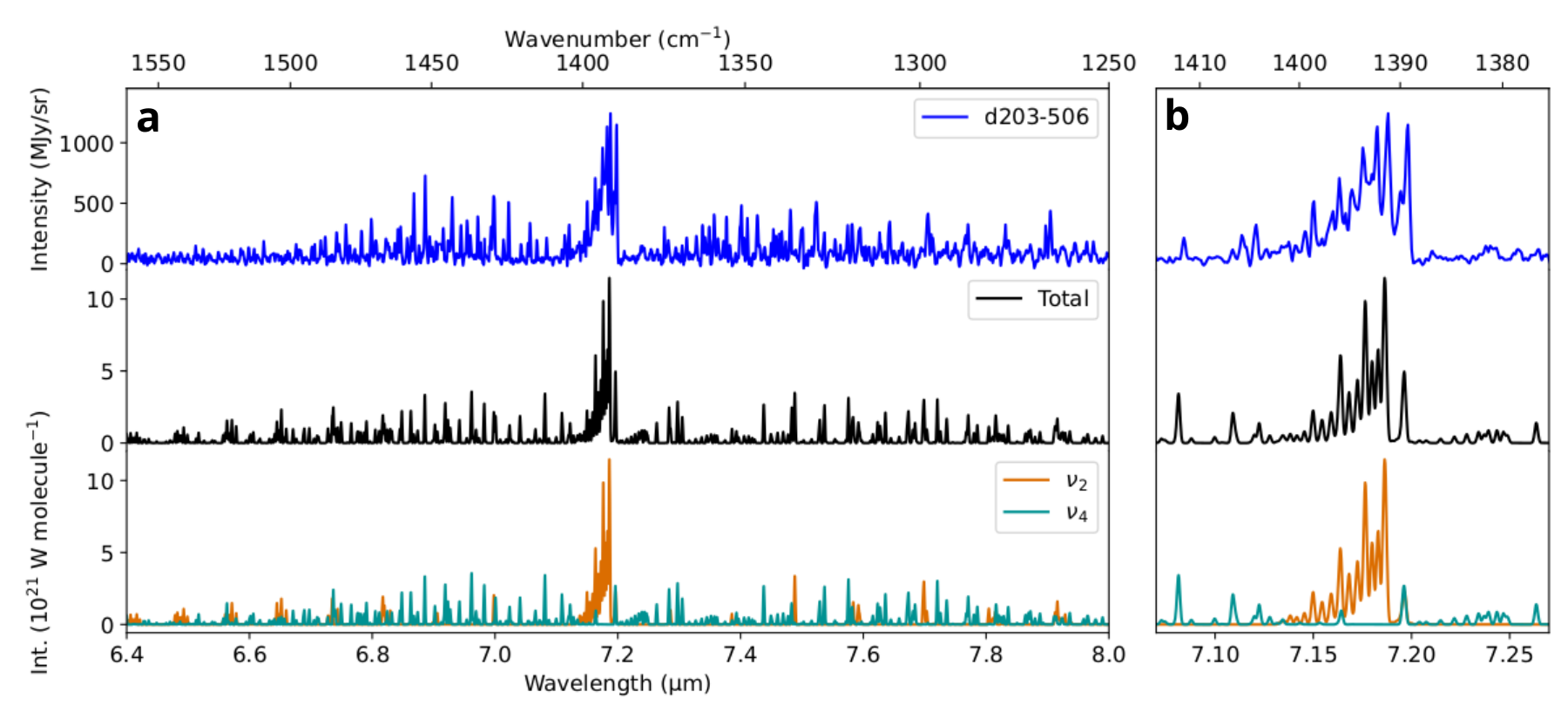}
    \includegraphics[width=\textwidth]{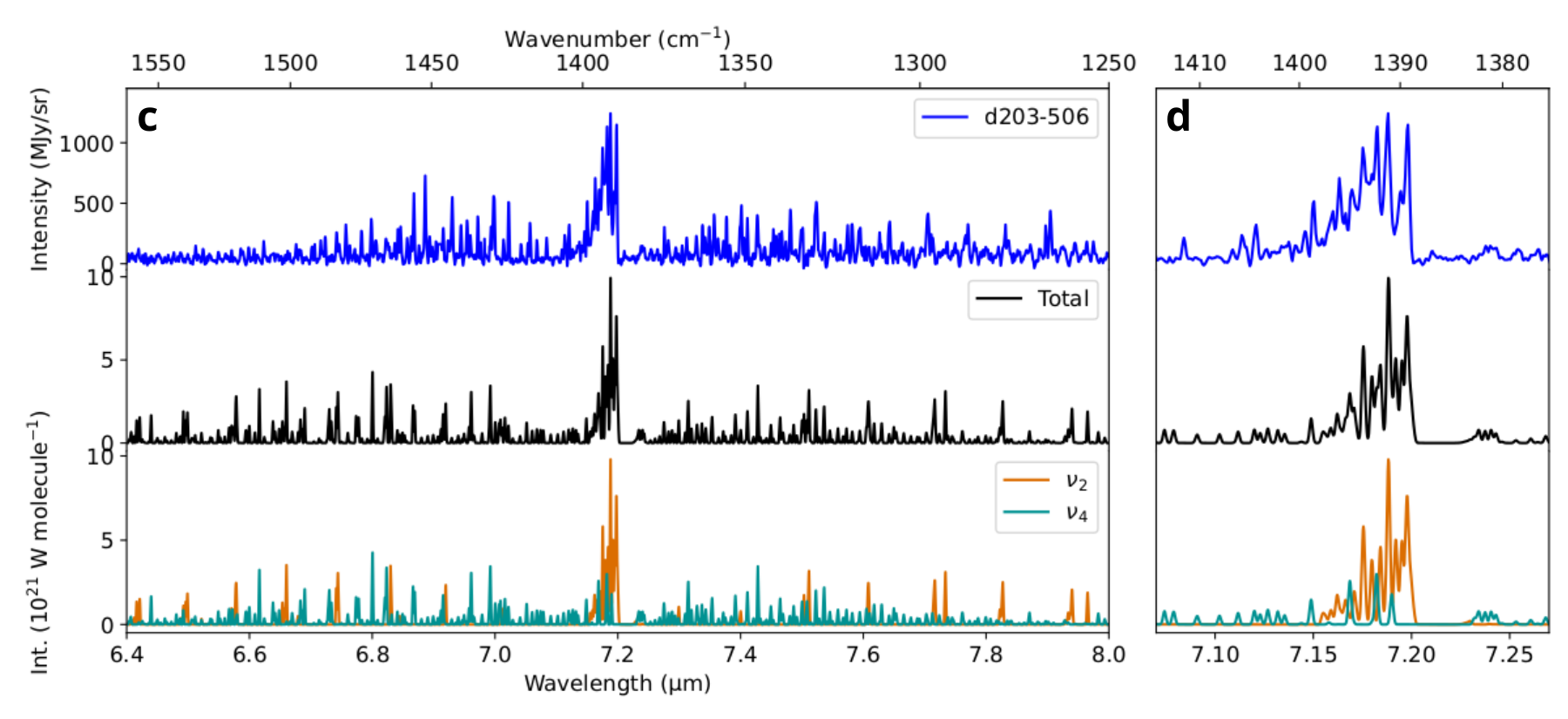}
    \caption{Model III. ({\bf a}), and Model IV. ({\bf b}) $T=400$ K, Gaussian profile (0.35 cm$^{-1}$ full-width-at-half-maximum)}
    \label{fig:modelIII}
\end{figure}

\begin{figure}[ht!]
    \centering
    \includegraphics[width=\textwidth]{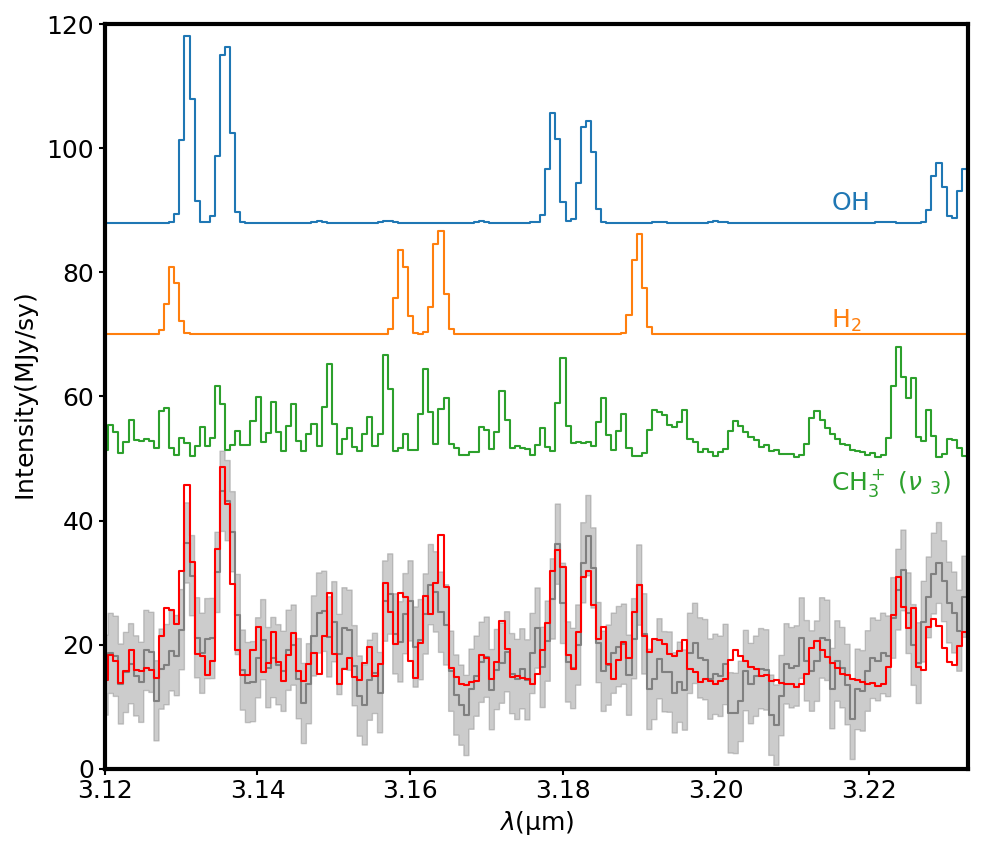}
    \caption{ { NIRSpec spectrum of d203-506. The spectrum is shown
    in gray, the shaded regions is the $\pm$ 3 sigma error interval of the data.  
    This includes the error provided by the JWST pipeline, and error 
    $\nu_3$ band of CH$_3^+$ in the NIRSpec spectrum of d203-506. Model of the OH emission (blue), H$_2$ emission (orange), CH$_3^+$ 
    emission (green), and sum of these three (red).  Beyond 3.22\um, emission due to the wings of the Aromatic Infrared Band at 3.3 \um is seen, affecting the baseline of the NIRSpec spectrum. The OH spectrum is computed with an LTE model at a temperature of 800K. A detailed model of the OH emission will be presented in a forthcoming paper (Zannese et al. in prep). The H$_2$ lines are fitted individually. The CH$_3^+$ model used here is Model III., at a temperature of 400K. }}
    \label{fig:NIRSpec}
\end{figure}

 \begin{figure*}
   \centering
   \includegraphics[width=\textwidth]{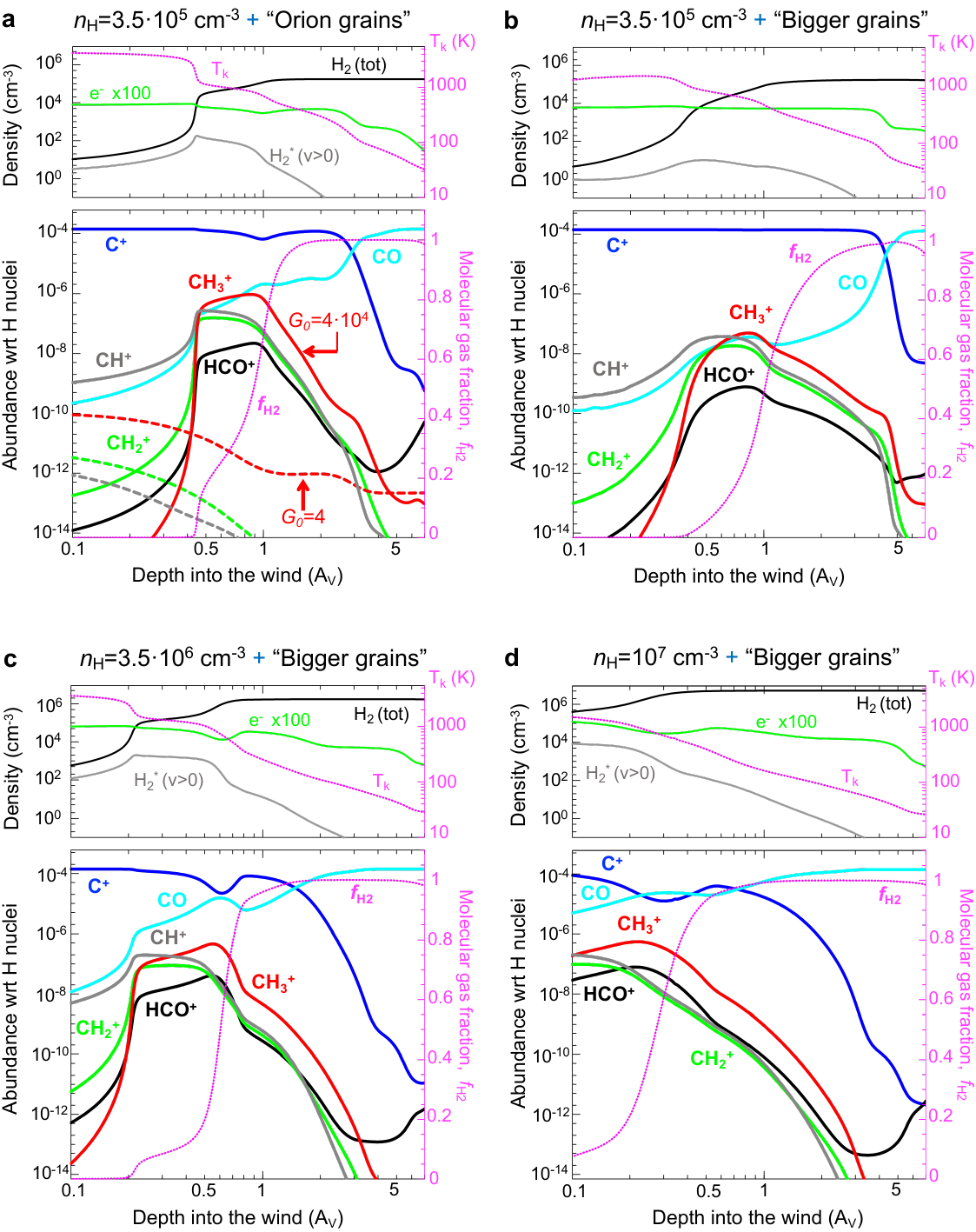}
      \caption{Photochemical model results for d203-506 adopting \mbox{$G_0$\,=\,4$\times$10$^4$} and different
       gas densities ($n_{\rm H}$) and dust grain properties.
       \textit{Upper panels}: 
       Density and gas temperature structure as a function of visual extinction ($A_V$) from the wind surface. The gray curve shows the density of vibrationally excited \mbox{H$_{2}^{*}$($v>0$)}.   
       \textit{Lower panels}: Abundance profiles with respect to H nuclei. The pink dotted curves show the molecular fraction $f_{H_2}$ profile. 
       Dashed curves in model a) refer to a model
with the same gas density but $G_0$ lower by a factor 10$^4$. }
         \label{fig:chemistry}
   \end{figure*}

 \begin{figure*}
   \centering
   \includegraphics[width=16cm]{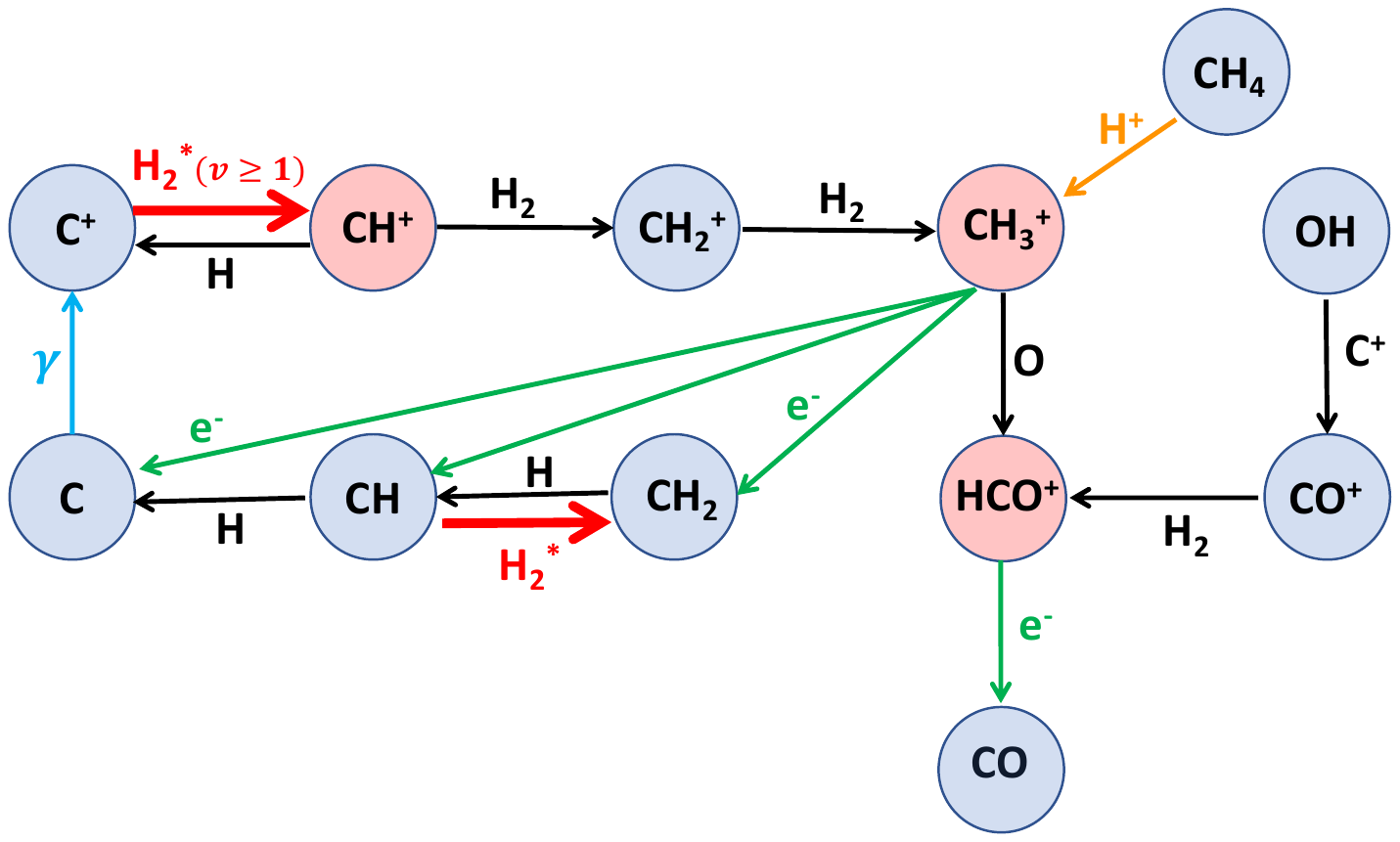}
      \caption{Dominant CH$_{3}^{+}$ formation and destruction reactions
      at the CH$_{3}^{+}$ abundance peak
      predicted by the  photochemical model shown in Fig.~\ref{fig:chemistry}.
      This reaction network also leads to abundant HCO$^+$  in \mbox{FUV-irradiated} gas layers where $x$(C$^+$)\,$>$\,$x$(CO). 
      Red arrows show endoergic reactions when H$_2$ is in the ground-vibrational
      state $v$\,=\,0. These reactions become fast only in disk layers where the gas temperature
    is high (several hundred K) and/or significant vibrationally excited 
    H$_{2}^{*}$\,($v$\,$\geq$0) exists. The formation of CH$_{3}^{+}$ from methane will only
    be relevant if very high CH$_{4}$ and H$^+$ abundances coexist in the  gas.}
         \label{fig:chemical_network}
   \end{figure*}

\end{document}